\documentclass[manuscript]{aastex}

\defcitealias{hel_sav_mom1999}{HSM99}
\defcitealias{ort_gra1988}{Ortolani \& Gratton 1988}
\defcitealias{mar_gal_apa1999}{MGA99}
\defcitealias{hol_smi_gri2000}{HSG00}
\defcitealias{can_flo1977}{Canterna \& Flower 1977}
\defcitealias{apa_hid2008}{AH08}

\begin{document}
\title{On the extended structure of the Phoenix dwarf galaxy}

\author{Sebastian L. Hidalgo\altaffilmark{1}}
\email{shidalgo@iac.es}
\and
\author{Antonio Aparicio\altaffilmark{1,2}}
\email{antapaj@iac.es}
\and
\author{David Mart\'{\i}nez-Delgado\altaffilmark{1}}
\email{ddelgado@iac.es}
\and
\author{Carme Gallart\altaffilmark{1}}
\email{carme@iac.es}

\altaffiltext{1}{Instituto de Astrof\'\i sica de Canarias. V\'\i a L\'actea s/n.
E38200 - La Laguna, Tenerife, Canary Islands, Spain.}
\altaffiltext{2}{Departamento de Astrof\'\i sica, Universidad de La Laguna/Instituto de
Astrof\'\i sica de Canarias. V\'\i a L\'actea s/n.
E38200 - La Laguna, Tenerife, Canary Islands, Spain.}

\begin{abstract}

We present the star formation history (SFH) and its variations with galactocentric distance for the Local Group dwarf galaxy of Phoenix. They have been derived from a (F555W, F814W) color-magnitude diagram (CMD) obtained from WFPC2@HST data which reaches the oldest main sequence turnoffs. The IAC-star and IAC-pop codes and the MinnIAC suite have been used to obtain the star formation rate as a function of time and metallicity, $\psi(t,z)$. We find that Phoenix has had ongoing but gradually decreasing star formation over nearly a Hubble time. The highest level of star formation occurred from the formation of the galaxy till 10.5 Gyr ago, when 50\% of the total star formation had already taken place. From that moment, star formation continues at a significant level until 6 Gyr ago (an additional 35\% of the stars are formed in this time interval), and at a very low level till the present time. The chemical enrichment law shows a trend of slowly increasing metallicity as a function of time till 8--6 Gyr ago, when Z starts to increase steeply to the current value.

We have paid particular attention to the study of the variations of the SFH as a function of radius. 
Young stars are found in the inner region of the galaxy only, but intermediate-age and old stars can be found at all galactocentric distances. The distribution of mass density in alive stars and its evolution with time has been studied. This study shows that star formation started at all galactocentric distances in Phoenix at an early epoch. If stars form {\it in situ} in Phoenix, the star formation onset took place all over the galaxy (up to a distance of about 400 pc from the center), but preferentially out of center regions. After that, our results are compatible with a scenario in which the star formation region envelope slowly shrinks as time goes on, possibly as a natural result of pressure support reduction as gas supply diminishes. As a consequence, star formation stopped first (about 7 to 8 Gyr ago) in outer regions and the scale-length of the stellar mass density distribution decreased with time. Finally, no traces of a true, old halo are apparent in Phoenix either in its stellar age distribution or in the stellar mass density distribution, at least out to 0.5 kpc (about 2.5 scale-lengths) from the center.

\end{abstract}


\section{Introduction}\label{secintroduccion}
During the 50's decade, Baade found an extended structure of red stars in a series of deep red plates of the IC 1613 dwarf galaxy. Designated as Baade's sheet \citep{san1971}, it was interpreted as being formed by Population II stars which extended by a factor of 2 with respect to Population I stars. During the last years, new extended structures have been found in several dwarf galaxies or have been confirmed:  IC 1613 \citep*{bat_dem_art2007}, WLM \citep{min_zij1996}, NGC 3109 \citep*{min_zij_alo1999}, Antlia \citep{apa_etal1997}, Phoenix \citep*{mar_gal_apa1999}, Leo A \citep{van_etal2004}, DDO 216 \citep{tik2006a}, DDO 190 \citep{apa_tik2000}, DDO 187 \citep*{apa_tik_kar2000}, NGC 6822 \citep*{kom_etal2003,bat_dem_kun2006}, DDO 165, and DDO 181 \citep*{hid_mar_apa2003}, to name but a few. Other authors have been preformed a systematic study of these extended structures in a wide sample of galaxies with the aim to derive global properties \citep*{dal_ber2002,tik2005,tik2006b,set_dal_dej2005a,set_dal_dej2005b}. The nature of these structures is still unclear. It is, however, of particular relevance to ascertain whether they are or not analogous of the halos of spiral galaxies; i.e. formed only by old stars (older than $\sim$ 10 Gyr) and with a hot kinematics. Within the context of the standard $\Lambda$-CDM, hierarchical scenario, large galaxy stellar halos of such characteristics are predicted to be formed as a consequence of the merging of dwarf building blocks. In principle, such a mechanism could also be at work in dwarf galaxies since, in the same framework, they are expected to be scaled-down versions of large galaxies \citep{moo_etal2001}.  

From the observational side, \citet{min_zij1996} and \citet{min_zij_alo1999} concluded that the extended structure they found in WLM is composed only by old stars and that it is a true halo analogous to those of spiral galaxies. However, from their analysis, based in color-magnitude diagrams (CMDs) that are not deep enough to reach the oldest turn-offs, it is not possible to exclude the presence of intermediate-age stars in the external regions of WLM. The same is true for other dwarf galaxies studied since then \citep{apa_etal1997, apa_tik_kar2000}. In addition, there are several independent observations that hint to the presence of intermediate-age populations in these low surface brightness envelopes of dwarf galaxies:  i) Carbon stars, which are bright members of the intermediate-age population, have been identified in dIrr galaxies, out to where young stars have already disappeared \citep*{bat_dem2000,dem_bat_art2006,bat_dem2006}; ii) PNe have been also detected at large distances from the center of many local dIrrs \citep{mag_etal2003,lei_etal2005}; iii) from population synthesis based on surface brightness fluctuations, \citet{hid_mar_apa2003} found indications that the extended regions of dwarf galaxies could contain a mixture of intermediate-age and old stars, the average age of such mixture possibly increasing with galactocentric distance;  iv) in irregular galaxies for which CMD reaching the oldest main-sequence turnoff exist, such as the Magellanic Clouds, an intermediate-age population has been found in the outskirts of the galaxy \citep{gal_etal2004b,noe_gal2007}.

A stellar 'halo' composed by both old and intermediate-age stars is difficult to reconcile with a true halo formed through the standard merging of smaller systems in a $\Lambda$-CDM framework: since most merging happens in the early history of the Universe, it is unlikely that merging produce significant intermediate-age stellar populations in halos. Alternatively, mixed-age 'halos' could be produced if stars form in the inner regions of the galaxy and migrate to the external ones as a consequence of velocity dispersion or heating produced by internal or external interactions, among other mechanisms. \citet{hid_mar_apa2003} proposed a 'shrinking' scenario in which the presence of intermediate-age populations in the outer parts of dIrr galaxies could be explained if a progressive shrinking of the star formation region, initially extending out to the galaxy outskirts, occurred in the galaxy, possibly produced by the gas progressively thinning out in the outermost regions. \citet{stin_etal2009} have shown, through SPH simulations, that a contracting star formation pattern results naturally as a consequence of gas pressure support reduction as gas supply is consumed in the center of the galaxy. Stellar migration from the center to the outer parts also plays a role, even though it is minor in general. 

Extended structures require a detailed study on an observational basis in order to ascertain their nature. First the stellar age and metallicity distributions as functions of the galactocentric distance must be determined. These distributions can be found by using deep CMDs (reaching the oldest turn-offs) to derive the star formation history (SFH). For galaxies beyond the Milky Way halo, the HST is required for this. Second, the kinematics in the inner and outer regions of the galaxies must be determined and compared with dynamical model predictions. Ground based, 10-meter class telescopes can reach the required accuracy for very nearby galaxies. 

This paper is devoted to the first topic, namely the age and metallicity distributions of the stellar population or, in other words, the star formation history (SFH), as a function of galactocentric distance in the Phoenix dwarf galaxy. Discovered by \citet{sch_wes1976}, it is a low surface brightness transition galaxy (dSph/dIrr) with $\rm \mu_B=24.9~ mag/('')^2$ \citep{mar_gal_apa1999}. At a distance of $\sim 450$ kpc from the Milky Way (\citealt*{hel_sav_mom1999}; \citealt*{hol_smi_gri2000}) and $\sim 600$ kpc form M31, Phoenix is close enough for observations to reach the old main sequence turn-offs but still far away enough to be unaffected by strong gravitational interaction with the Milky Way or M31. Phoenix contains a young stellar population with strong evidence of a population gradient as seen from ground-based observations \citep{ort_gra1988,hel_sav_mom1999,hol_smi_gri2000}. \citet{mar_gal_apa1999} and \citet{hel_sav_mom1999} found that the younger and bluer stars of Phoenix are concentrated in the center of the galaxy. The same authors also found that RGB stars are present throughout the whole body of Phoenix. Associated with this stellar population gradient, \citet{mar_gal_apa1999} found a sharp rotation in the elliptical isopleth of $\sim 90^\circ$ at radius $r \sim 115\arcsec$ from the center, suggesting the existence of two components: an inner one elongated in the east-west direction, which contains all the young stars, and an outer component oriented north-south, which lacks young stars. Limited by the depth of their photometry, they were unable to determine the nature of the stellar populations in the outer component. Phoenix has no \ion{H}{1} emission coincident with its main body which suggest that all the gas of the galaxy has been removed or exhausted by the star formation processes. There are some \ion{H}{1} clouds detected between $4.5\arcmin$ and $9\arcmin$ from the galaxy center \citep*{car_dem_cot1991,you_lo1997,you_etal2007,hel2002} that seem to be related to Phoenix \citep{gal_etal2001,irw_tol2002}. \citet{you_etal2007} have used high sensitivity and resolution imaging with the VLA to characterize the morphology of the gas cloud and its mass. The cloud, with a mass $M_{HI} \simeq 10^5 M_{\odot}$, has a crescent-like shape with the center of curvature in the direction of the galaxy, which implies that the gas was likely expelled from the galaxy by supernovae winds.   These characteristics make Phoenix a case of special interest both to characterize in detail its full SFH  and its variation with galactocentric radius. This can provide important insight on the nature of the extended structures in dwarf galaxies. Main properties of Phoenix are summarized in Table \ref{phoenixprop}.

The organization of this paper is as follows: In \S\ref{secreduccion} the data, the photometric procedures, and the calibration method are presented. In \S\ref{seccmd} the CMD is presented. In \S\ref{secdistancia} the distance to the galaxy is obtained using the tip of the Red Giant Branch (TRGB). In \S\ref{seccumulos} three former candidates to globular clusters are discussed. Section \ref{seccompletitud} describes the completeness tests performed. The IAC-star \citep{apa_gal2004} and IAC-pop \citep{apa_hid2009} algorithms, together with the suite MinnIAC (see also \citealt{apa_hid2009}) have been used to obtain the SFH. It is presented in \S\ref{secsfh} for the whole covered field and in \S\ref{sfhradio} as a function of galactocentric radius. The evolution of the galaxy structure with time is discussed in \S\ref{evogal}. Finally, the main conclusions derived in this paper are summarized in \S\ref{concl}.

\section{Reduction and calibration of data}\label{secreduccion}

The images of Phoenix were obtained in January 2001 using the WFPC2 aboard the HST (cycle 9, proposal ID 8769, P.I. A. Aparicio). The WFPC2 has four $\rm 800\times 800~ pixel$ CCDs; three of them (the Wide Field Camera, WFC) with a pixel size of $\rm 0.1''~pixel^{-1}$ and the remaining one (the Planetary Camera, PC) with a pixel size of $\rm 0.046''~pixel^{-1}$. The resulting field of view are $150\times 150~('')^2$ for the WFC and $35\times 35~('')^2$ for the PC. Images were taken in two fields observed in the HST filters F555W and F814W: an inner field centered on the main body of the galaxy (observed on 2001 Jan 10 and 11), and another located $\sim 2.7'$ from the former (observed on 2001 Jan 16 and 17).

\begin{figure}
\centering
\includegraphics[width=10cm,angle=0]{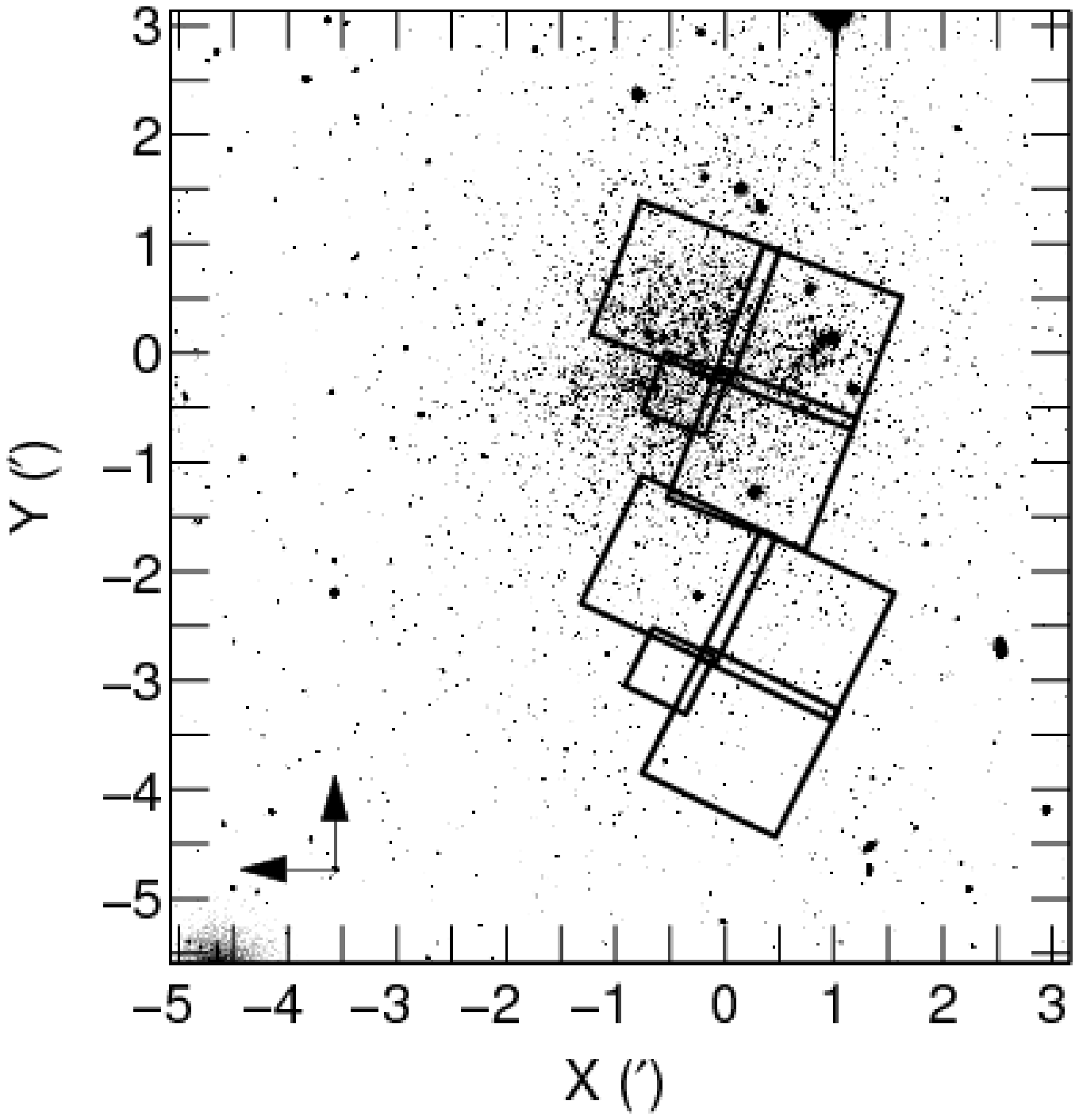}
\protect\caption[ ]{Image of Phoenix obtained with the 100-inch {\it Du Pont} telescope at Las Campanas Observatory (Chile). The Fields observed with HST are overplotted on the image. The center of the figure is located at the optical center of the galaxy. At the distance of Phoenix, 1\arcmin~ corresponds to $\rm \sim 122~ pc$. North is up, East is to the left. 
\label{f01}}
\end{figure}

Images obtained by \citet*{smi_gri_hol1997} (cycle 6, proposal ID 6798, P.I. G. Smith), which match our inner field described above, were also used. Total integration times including  \citet{smi_gri_hol1997} data, were 12100 s and 14500 s for the F555W and F814W filters respectively for the inner field; and 11900 s and 14300 s for the F555W and F814W filters respectively for the outer field. The observation log is given in Table \ref{orbphoenix}. Figure \ref{f01} shows the observed fields overplotted on an image of the galaxy obtained with the 100-inch {\it Du Pont} telescope at Las Campanas Observatory (Chile). 

\notetoeditor{PLEASE, PUT TABLE \ref{orbphoenix} HERE}

\notetoeditor{PLEASE, PUT FIGURE \ref{f01} HERE}

Processed images of Phoenix were retrieved from the HST archive. There is a small difference in the pixel size of the WFPC2 CCDs. Therefore, flux correction per pixel was necessary using an image mask kindly provided by P. B. Stetson. Bad pixels were corrected with another image mask. The process to obtain the photometry follows the method described by \citet{tur1998} and is based on the application of DAOPHOT and ALLFRAME \citep{ste1994}. Moffatian PSFs of parameter $\beta=1.5$ and 20 pixel radius, provided by P. B. Stetson, were used.

Photometry was corrected for the charge transfer efficiency (CTE) effect. The correction depends on the position of the stars in the CCD, its temperature, and the observation date. CTE correction must be applied to the total magnitude of the stars, i.e., after aperture corrections. The aperture corrections proposed by \citet{hil_etal1998} were used. The difference between the focus of the HST for the data of \citet{hil_etal1998} and the data used here is $4~\mu m$, which corresponds to a difference in the flux of the stars of less than 0.5\%. This small difference lends support for us to assume the correction apertures of \citet{hil_etal1998} to estimate the CTE corrections for our photometry.

\begin{figure}
\centering
\includegraphics[width=10cm,angle=0]{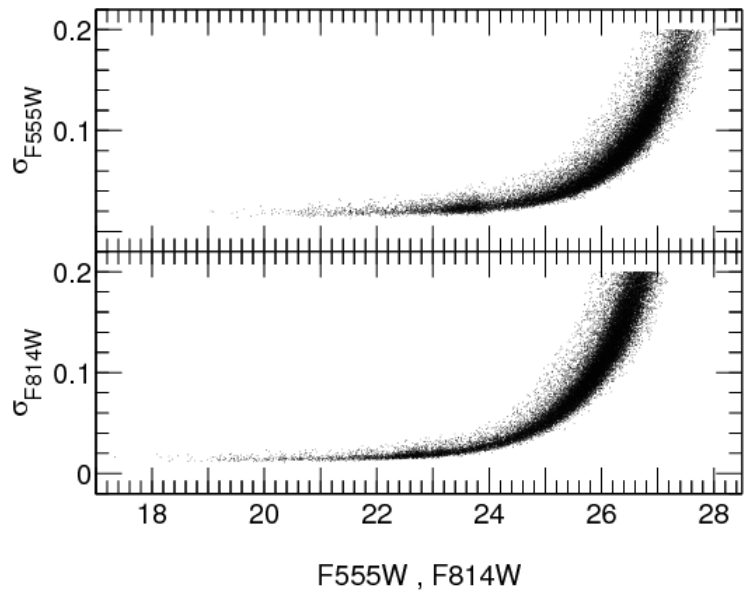}
\protect\caption[ ]{Uncertainties provided by ALLFRAME as a function of magnitude for F555W (top panel) and F814W (bottom panel) filters.\label{f02}}
\end{figure}

The final photometric list was cleaned of non-stellar objects and badly measured stars by limiting the $\sigma$ and SHARP values provided by ALLFRAME ($\sigma\leq 0.2$, $-0.5\leq SHARP\leq 0.5$). No limits in {\it CHI} (also provided by ALLFRAME) value were imposed because of the restrictions in $\sigma$ eliminated those stars with large {\it CHI}. Figure \ref{f02} shows the errors provided by ALLFRAME as a function of magnitude.

\notetoeditor{PLEASE, PUT FIGURE \ref{f02} HERE}

The photometric calibration was carried out using $V$ and $I$ ground based observations of Phoenix from the 100-inch {\it Du Pont} telescope at Las Campanas Observatory (Chile) \citep{gal_etal2004b}. A selection of $\sim 1600$ bright and isolated stars in Phoenix observed from the ground were used as local standard stars. The photometric transformation equations were obtained comparing the magnitudes of local standards in $V$ and $I$ Johnson-Cousins filters with the magnitudes of the same stars observed in the HST F555W and F814W filters. For stars of spectral types O5V to M6V, these filters are very close approximations to the Johnson-Cousins filter $V$ and $I_c$ respectively \citep{bag_etal2002}. The WFPC2 calibration was done independently for each CCD. The final photometric calibration equations can be expressed as:

\begin{eqnarray}
V-v & = &Z_V + C_V\cdot(V - I) + K_V\cdot Y\nonumber\\
I-i & = &Z_I + C_I\cdot(V - I) + K_I\cdot Y,\nonumber\\
\end{eqnarray}

where $V$ and $I$ stand for calibrated magnitudes and $v$ and $i$ are the instrumental magnitudes. The coefficients $Z_{V,I}$  and $C_{V,I}$ are, respectively, the zero point of the photometry and the color term. The $K_{V,I}$ coefficient is a term that accounts for the dependence of the magnitude with the Y coordinate of the star on the CCD. The correction introduced by this term makes the magnitude of a star dimmer with increasing Y (CTE has the opposite effect). This could be either a new term in the photometric calibration of the WFPC2 or a correction to the CTE. Table \ref{calphoenix} lists the values of the coefficients for each CCD and filter (column 1). The photometric zero point ($Z$) is given in column 2. Column 3 gives the color term ($C$). The term $K$ for the dependency on the $Y$ coordinate is given in column 4. Finally, the photometric transformation error is given in column 5. There are two sources contributing to the latter: i) the error of the ground based photometric transformations, which is the dominant, and ii) the error of the transformation from the HST to ground based photometry. Both have been estimated as the rms of the corresponding data distributions divided by $\sqrt{N-2}$, where $N$ is the total number of measured stars in each case. The error given in column 5 of Table \ref{calphoenix} is the quadratic sum of both.

\notetoeditor{PLEASE, PUT TABLE \ref{calphoenix} HERE}

\section{The CMD of Phoenix}\label{seccmd}

\begin{figure}
\centering
\includegraphics[width=10cm,height=10cm,angle=0]{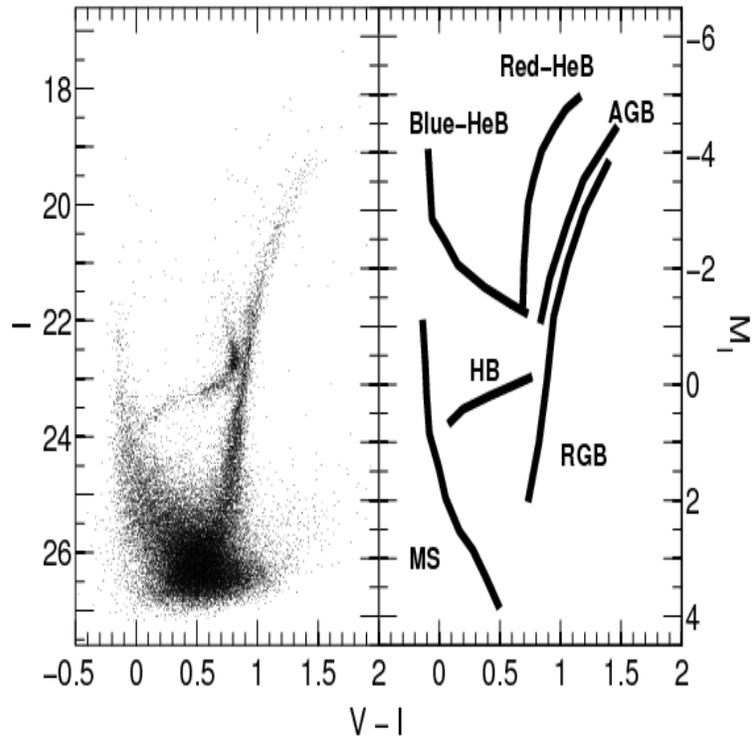}
\protect\caption[ ]{CMD of Phoenix. Absolute magnitudes are shown in the right axis. Main features of the CMD are shown on the right panel.\label{f03}}
\end{figure}

The calibrated CMD of Phoenix is shown in Fig. \ref{f03}. Absolute magnitudes are given on the right by using a distance modulus of $(m - M)_0 = 23.09$ mag and extinctions $A_V=0.062$ mag and $A_I=0.036$ (see \S\ref{secdistancia}). The CMD shows a clear main sequence (MS) and RGB. The brightest stars in the MS are younger than 1 Gyr, while the RGB is formed of stars older than $1-2$ Gyr. Parallel to the blue side of the RGB, some AGB stars can be observed. A few stars above the TRGB could be intermediate-age AGB stars. An extended Horizontal Branch (HB) is also present. This is an indication of very old, metal poor stars. The gap produced in the HB by the RR-Lyrae variables is also visible. Near the intersection of the HB with the RGB, the Red Clump (RC) is observed. Finally, some blue core helium-burning (Blue-HeB) and red core helium-burning (Red-HeB) stars are seen extending to brighter magnitudes above the RC. These stars are younger than $\rm \sim 0.5~ Gyr$. In short, the CMD of Phoenix shows stars of all ages.

\notetoeditor{PLEASE, PUT FIGURE \ref{f03} HERE}

\section{Distance to Phoenix}\label{secdistancia}

\begin{figure}
\centering
\includegraphics[width=10cm,angle=0]{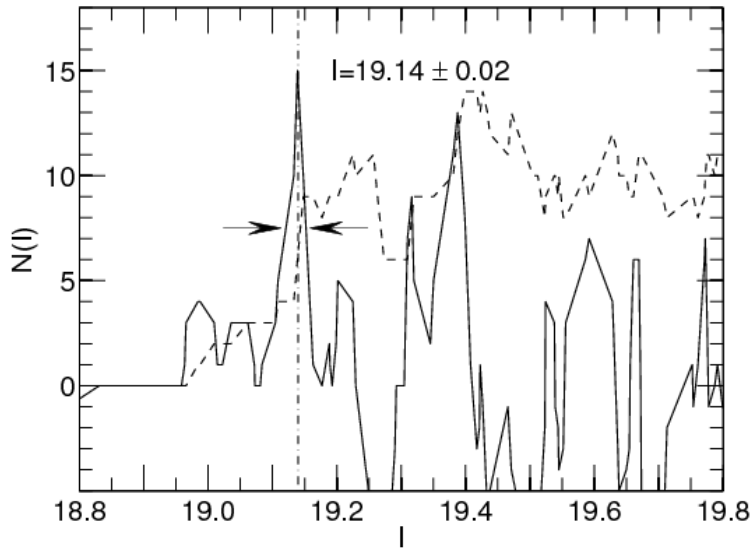}
\protect\caption[ ]{The luminosity function of stars redder than $(V-I)=0.8$ (dashed line) and its convolution with a Sobel filter of kernel [1,2,0,-2,-1] (solid black line). A vertical dash dotted line shows the convolution peak, placed at $I=19.14\pm 0.02$ mag. Arrows indicate the peak half height. The error of the peak position is estimated as the width at half of height of the peak.\label{f04}}
\end{figure}

The distance to Phoenix has been estimated from the TRGB magnitude, following the prescriptions method by \citet{bel_etal2004}. For our determination of the TRGB, a Sobel filter of kernel [1,2,0,-2,-1] \citep{mad_fre1995} was convolved with the luminosity function of the RGB stars redder than $(V - I) = 0.8$ with magnitudes $18.5\leq I\leq 21$. The luminosity function was obtained using a binning of 0.1 mag with a step of 0.05 mag. Figure \ref{f04} shows the result of the convolution. The TRGB is found to be at $I_{TRGB}=19.14\pm 0.02~\rm mag$. Using extinctions $A_V=0.062$ mag and $A_I=0.036$ mag \citep*{car_cla_mat1989}, the TRGB corrected of reddening is found to be at $I_{TRGB,0}=19.10\pm 0.02~\rm mag$ which is, within the errors, located at the same magnitude found by \citet{hol_smi_gri2000}. The error was estimated as the half maximum of height of the convolution peak.

The absolute magnitude of the TRGB has been calculated using the results by \citet{bel_etal2004} and assuming [M/H]=[Fe/H]. The metallicity [Fe/H] has been estimated using the mean color of the RGB stars at $I_0=-3.5$ and the empirical calibration given by \citet{sav_etal2000}. The RGB mean color is $(V - I)_{-3.5,0}=1.32\pm 0.01$, which gives $\rm [Fe/H]=-1.87\pm 0.06~dex$ ($Z=(2.6\pm 0.3)\times 10^{-4}$). Using  $\rm [Fe/H]=-1.87\pm$, we obtain an absolute magnitude of the TRGB $M^{TRGB}_{I,0}=-3.99$ mag which results in a distance modulus of $(m - M)_0 = 23.09\pm 0.10$, or a distance $d=415\pm 19$ kpc. The errors include the uncertainties in the calibration method and the determination of the magnitude of the TRGB.

\notetoeditor{PLEASE, PUT FIGURES \ref{f04} HERE}

The distance obtained here is in full agreement with the distances obtained by \citet*{ryd_dem_kun1991} ($(m-M)_0=23.1\pm 0.1$), by \citet{mar_gal_apa1999} ($(m-M)_0=23.0\pm 0.1$), by \citet{hel_sav_mom1999}  ($(m-M)_0=23.21\pm 0.08$), by \citet{hol_smi_gri2000} ($(m-M)_0=23.11$), and with the most recent distance estimation by \citet{men_etal2008} using infrared photometry and variable stars ($(m-M)_0=23.1\pm 0.2$). The obtained metallicity ($\rm [Fe/H]=-1.87\pm 0.06~ dex$) is also in good agreement with those obtained by \citet{hel_sav_mom1999} ($\rm [Fe/H]=-1.81\pm 0.10~ dex$), and by \citet{hol_smi_gri2000} ($\rm [Fe/H]=-1.81~ dex$), but it is lower than the metallicity obtained by \citet{mar_gal_apa1999} ($\rm [Fe/H]=-1.37~dex$) using ground based observations. This discrepancy for the metallicity could be explained by a mistake in the sign of one of the color terms in the equations used by \citet{mar_gal_apa1999} to transform the instrumental magnitudes into the Johnson-Cousins system \citep[see][]{gal_etal2004b} which affects the color of the RGB.

\section{The globular clusters candidates in Phoenix}\label{seccumulos}

Globular clusters (GC) detected in the Local Group show a luminosity distribution in the range of $M_V\geq-7.1$ and $0.7<(V-I)_0<1.9$ \citep{har1991}. They are usually selected from images in which they are unresolved using these criteria as well as the PSF profiles. However, these characteristics may also correspond to background galaxies.

\begin{figure}
\centering
\includegraphics[width=10cm,angle=270]{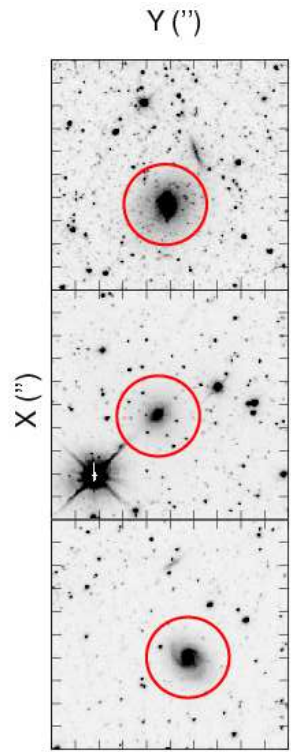}
\protect\caption[ ]{Globular cluster candidates {\it ''a'', ''b'', {\rm and} ''c''} (from top to bottom) proposed by \citet{mar_gal_apa1999}. All of them seem to be background galaxies.
\label{f05}}
\end{figure}

There are currently six globular clusters candidates in Phoenix identified in previous works \citep{can_flo1977,mar_gal_apa1999}.  We have identified in our HST images the three objects proposed by \citet{mar_gal_apa1999}. They are shown in Figure \ref{f05}. Visual inspection indicates that the objects are background galaxies. Candidates identified by \citet{can_flo1977} are out of our observed field.

\notetoeditor{PLEASE, PUT FIGURE \ref{f05} HERE}

\section{Completeness tests}\label{seccompletitud}

Our analysis of the photometry completeness follows the procedure proposed by \citet{apa_gal1995}. In short, a list of artificial stars is made. The absolute magnitudes of these stars are transformed to instrumental magnitudes with the equations used to calibrate the instrumental photometry  (CTE, aperture corrections, zero points and color terms) and assuming the distance modulus and reddening given in \S\ref{secdistancia}. The individual PSF for each frame is used to inject the artificial stars into the single frames. 

A synthetic CMD with 500000 stars was used as catalog of artificial stars. It was generated with IAC-star \citep{apa_gal2004} using the BaSTI \citep{pie_etal2004} stellar evolution library and the bolometric corrections by \citet{cas_kur2004}. A constant star formation rate (SFR) between 0 and 14 Gyr was assumed. An uniform range of metallicity $0.0001\leq Z\leq 0.002$ for all ages was used. The criterion in the selection of the IAC-star input parameters is that colors, magnitudes, ages, and metallicities of the artificial stars cover the ranges shown or reasonably expected for the observed stars. This is important in order of properly simulate observational effects on other synthetic CMDs. The artificial stars were distributed over the 8 CCD chips, the number of injected stars in each field being proportional to the number of observed stars. They were distributed in a triangular grid with a separation between star centers of $2\times R_{psf} + 1$ pixels, where $R_{psf}$ stands for the PSF radius. This separation prevents the superposition of artificial stars to each other.

Once the artificial stars were injected, the new synthetic frames were processed photometrically in the same way and with the same parameters as the original images. The same photometric calibration corrections were applied to the magnitudes of the fake recovered stars, with the exception of the CTE. Since this is a process that affects the read out of the CCD, it is not expected to affect the artificial stars, so we did not include it in the photometric corrections for these stars. 

\begin{figure}
\centering
\includegraphics[width=12cm,height=12cm,angle=0]{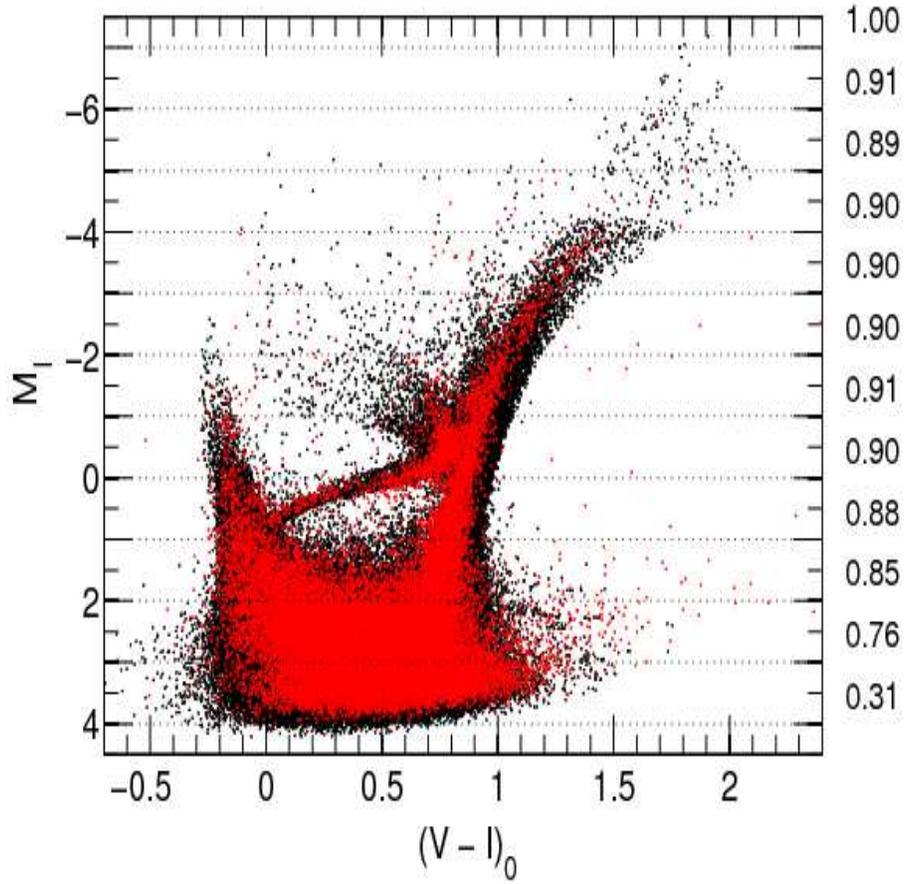}
\protect\caption[ ]{Observed CMD (red) overplotted on the CMD of the recovered artificial star CMD (black). Completeness is given on the right Y-axis \label{f06}}
\end{figure}

Figure \ref{f06} shows the CMD of Phoenix overplotted on the recovered artificial star CMD. The completeness factors are shown on the right $Y$-axis as a function of magnitude. It can be observed that the recovered artificial star CMD fully covers the observed CMD, fulfilling our previous requirement.

\notetoeditor{PLEASE, PUT FIGURE \ref{f06} HERE}

\section{The Star Formation History of Phoenix}\label{secsfh}

We have obtained the SFH of Phoenix following the method explained by \citet{apa_hid2009}. It makes use of the suite of routines MinnIAC and the algorithms IAC-pop (see \citealt{apa_hid2009} for both) and IAC-star \citep{apa_gal2004}. The most important characteristics of our method (see below and \citet{apa_hid2009} for a more comprehensive discussion) compared with others SFH recovering methods as MATCH \citep{dol1997,dol2002} or Star-FISH \citep{har_zar2001} are: i) interpolation in both age and metallicity directly on stellar evolution tracks to produce synthetic CMD with smooth sampling of both coordinates; ii) possibility to use several sets of stellar evolution models (currently \citet{ber_etal1994,pie_etal2004,gir_etal2000}); iii) the powerful minimization technique implemented in IAC-pop (which makes use of a genetic algorithm \citep{cha1995}), allows us to solve for a huge number of parameters (e.g. well sampled age-metallicity relation); iv) different areas of the synthetic and observed CMDs are sampled with different resolution, according to the reliability of our theoretical knowledge of these areas; v) each SFH solution is given as the average of a number of solutions obtained from multiple samplings of both a) the synthetic CMD in terms of simple stellar populations and b) the CMDs in terms of boxes in the color-magnitude plane.

We define the SFH as a distribution function $\psi(t,z){\rm d} t{\rm d} z$ that can be identified with the usual SFR but with dependence both on time and metallicity. An arbitrary $\psi(t,z)$ can be written as a linear combination of the SFRs, $\psi_i$, of simple stellar populations, each containing stars with ages and metallicities within small intervals:

\begin{equation}
\psi(t,z)=A\sum _i\alpha_i\psi_i
\end{equation}

\noindent where the SFR is constant within each $\psi_i$ and $A$ is a scaling constant. 

IAC-pop derives $\psi(t,z)$ of the target object by comparing the star distributions in the observational CMD (oCMD) and in a synthetic CMD. A detailed description of the method is given in \citet{apa_hid2009}. In short, the procedure is as follows. A large synthetic stellar population with ages and metallicities spanning ranges wider than those expected for the target object is computed together with its CMD. We denote this starting global CMD as sCMD. This population is divided in a set of simple stellar populations, $\psi_i$. Each one has an associated CMD, which distribution of stars as a function of color and magnitude is parametrized using a combination of grids (see below and \citealt{apa_hid2009}). The CMD parametrization of simple population $i$ is denoted as $M_i^j$, where $j$ refers to the $j$-th grid element. Using this information, the parametrization of the star distribution in the synthetic CMD of any SFH that can be written according to Eq. (2), can be obtained as $M^j=A\sum _i\alpha_iM_i^j$. It should be noted that no need of explicit computation of the new synthetic CMD is required for this and that it is enough to work with the $M_i^j$ counts. A similar parametrization, denoted $O^i$, is done for the observational CMD. At this point, the $\alpha_i$ coefficients of the SFH of the target object are obtained by minimizing a merit function comparing $M_i^j$ and $O^j$. Mighell's $\chi^2_\gamma$ \citep{mig1999} is used for this purpose. 

Concerning the software, IAC-star is used for the computation of synthetic populations and their CMDs; IAC-pop computes the $M^j$ distributions and search for the solutions, and MinnIAC takes care of the several CMD parametrization as well as the solution handling and error computations (see below).

\begin{figure}
\centering
\includegraphics[width=12cm,angle=0]{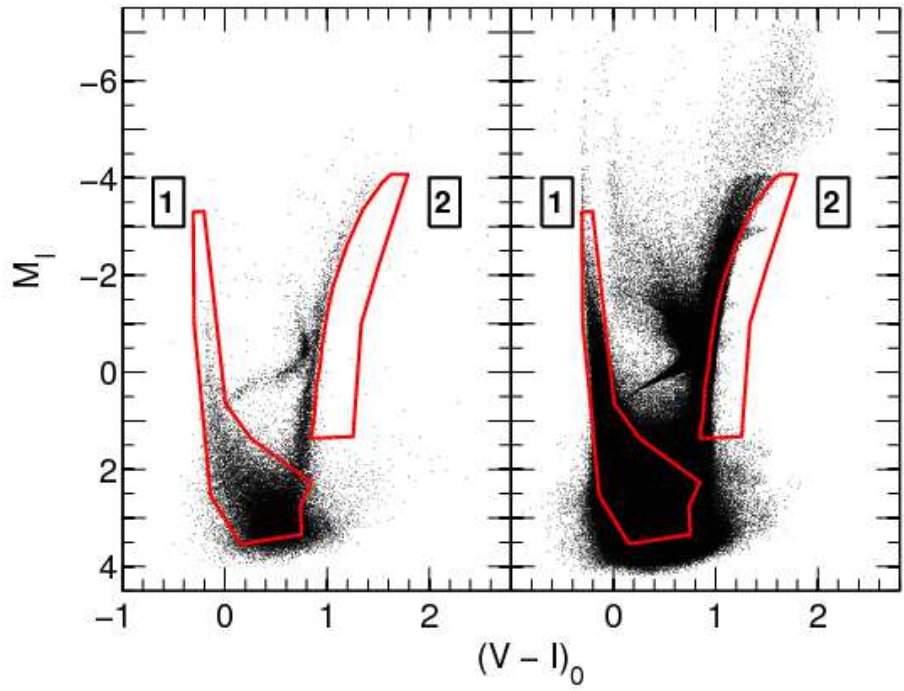}
\protect\caption[ ]{Regions of the CMDs used to obtain the SFH. The observed CMD is shown on the left and the sCMD, on the right. 
\label{f07}}
\end{figure}

Specifically, the steps followed to obtain the SFH of Phoenix were as follow:

\begin{itemize}

\item The BaSTI \citep{pie_etal2004} stellar evolution library was used to compute the sCMD. It contains $5\times 10^6$ stars with a constant SFR between 0 and 13.5 Gyr and an uniform range of metallicity $0.0001\le Z\le 0.002$ for all ages. These limits were selected to cover the full expected ranges of stellar ages and metallicities. Six fractions of binary stars were considered: 0, 20, 40, 60, 80 and 100\% with mass ratio randomly distributed between 0.5 and 1. Bolometric corrections by \citet{cas_kur2004} were used in all the cases and the Initial Mass Function (IMF) was that by \citet*{kro_tou_gil1993}. 

\item Observational effects were simulated in all sCMD using the results of the experiments described in \S\ref{seccompletitud}. The simulation is done as follows: for each synthetic star in sCMD with magnitude $m_s$, a list of artificial stars with $|m_i - m_s|\leq \epsilon$ is selected for each filter, being $\epsilon$ an input interval. $\epsilon=0.2$, which is the largest error in the photometry, has been used. A single artificial star is randomly selected. If $m_i^\prime$ and $m_r^\prime$ are the injected and recovered magnitudes of the selected artificial star in a given filter, then $m_s^e=m_s + m_i^\prime - m_r^\prime$ will be the magnitude of the synthetic star with observational effects simulated. The same is done for both filters. With this procedure, the observational effects are simulated in sCMD star by star. The quality of the observational effects simulations is determined by the number of injected stars in the completeness tests in comparison with the number of synthetic stars in sCMD.

\item Simple stellar populations were selected from sCMD by using a 0.5 Gyr age binning for stars with ages $0-1$ Gyr; 1 Gyr for stars within $1-6$ Gyr years old, and 2 Gyr for stars older than 6 Gyr. The metallicity bin size increase with metallicity: $\Delta z=0.0001$ for stars with $z\leq 0.0005$; $\Delta z=0.0002$ for stars with $0.0005< z\leq 0.0007$; $\Delta z=0.0003$ for stars with $0.0007< z\leq 0.001$; and $\Delta z=0.0005$ for stars with $z>0.001$. 

\item According to \citet{apa_hid2009}, an optimum parametrization of the CMD stellar distribution for the purpose of deriving the SFH is obtained defining several significant regions (or {\it bundles}) in the CMD and sampling them with grids which bin size is fixed for each bundle. In our case, two bundles have been defined (see Fig. \ref{f07}); one of them on the MS and sub-giant region and the other one on the reddest part of the RGB. The former includes $\sim 65\%$ of the observed stars and was sampled by a grid with boxes of size $0.015\times 0.4$ in color and magnitude. The latter contains only $\sim 100$ observed stars and was sampled with boxes of size $C\times 0.4$, with $C$ covering the full color interval of the region for each magnitude.

\notetoeditor{PLEASE, PUT FIGURE \ref{f07} HERE}

\item To minimize the dependency of the SFH on the selection of the simple stellar populations and position of the sampling grids, MinnIAC generates several sets of grids to parametrize the CMDs and several sets of simple stellar populations. This is done by introducing offsets in the positions of the CMDs sampling boxes and by moving the limits $b$ of the age and metallicity bins defining the simple stellar populations to $b\pm 0.25\,\delta b$ where $\delta b$ is the bin size.

\item Uncertainties in distance, reddening and photometric zero-point may have significant effects on the SFH solution. To overcome this, several offsets have been applied to the observational CMD and the SFH has been found for each of them. Offsets were applied according to a grid of $5\times 5$ points separated by intervals of 0.02 in $(V-I)$ and 0.05 in $I$. 

\item In summary, for each binary fraction and CMD zero-point offset, the different sets of simple stellar populations and boxes defined above give a total of 16 parametrisation of the CMDs (8 sets of simple stellar populations $\times$ 2 grid offsets). The average of the 16 solutions is assumed as the final solution for the corresponding binary fraction and CMD zero-set offset, and the rms dispersion is assumed to be the associated error (see \citealt{apa_hid2009}). The average, $\chi^2_{av}$, of the 16 $\chi^2$ is also computed and considered an indicator of the goodness of the average solution. The solution for the binary fraction and CMD zero-point off-set producing the smallest $\chi^2_{av}$ is assumed to be the solution for the SFH of Phoenix. 

\end{itemize}

The final solution has $\chi^2_{av}=1.9$ and corresponds to 40\% of binaries and CMD offsets of $\Delta (V-I)=0.03$ and $\Delta I=0.075$. It is interesting to note that the solution (and its $\chi^2_{av}$) does not change significantly for binary fractions in the range of 40\% to 80\%. Also, the $\Delta (V-I)$ and $\Delta I$ found should not be used to correct distance or reddening, since they are also accounting for possible errors in the stellar evolution library and in the photometry zero-point.

\begin{figure}
\centering
\includegraphics[width=12cm,angle=0]{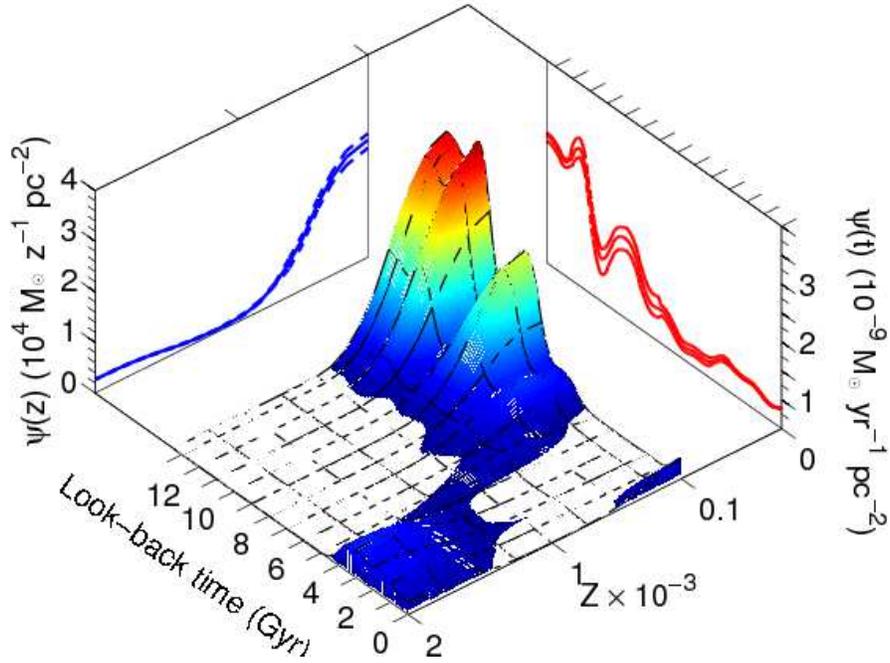}
\protect\caption[ ]{The SFH $\psi(t,z)$ of Phoenix. It contains a binary fraction of 40\%. The height of the curved surface over the age--metallicity plane gives $\psi(t,z)$. The SFH depending only on time ($\psi(t)$) and metallicity ($\psi(z)$), are shown in the right and left vertical pannels, respectively. The CEL is the projection onto the horizontal plane.
\label{f08}}
\end{figure}

The population box \citep{hod1989} for the SFH is shown in Fig. \ref{f08}. The height of the curved surface over the age--metallicity plane gives $\psi(t,z)$. The total mass converted into stars per unit area ($M_{\odot}~ pc^{-2}$) in the age--metallicity interval ($[t_i,t_i+\delta t]$;$[z_j,z_j+\delta z]$) is the volume below the solution and within that interval: $m_{ij}=\int_{t_i}^{t_i+\delta t}\int_{z_j}^{z_j+\delta z}\psi(t,z){\rm d} t{\rm d} z$. The SFH as a function of time $\psi(t)$ and metallicity $\psi(z)$ only are also given as projections of $\psi(t,z)$.

\begin{figure}
\centering
\includegraphics[width=12cm,angle=0]{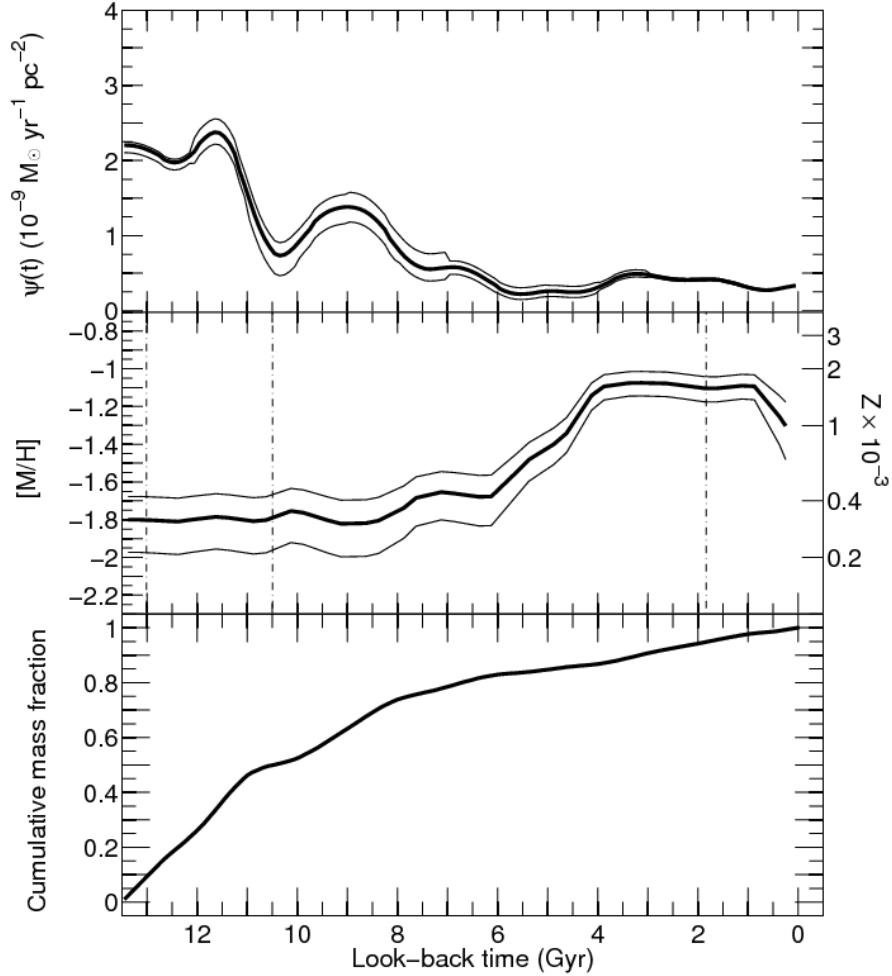}
\protect\caption[ ]{Upper panel shows the SFH of Phoenix as a function of time only, $\psi(t)$. It is the same shown in the right-hand vertical panel of Fig. \ref{f08}. Thin lines represent the rms dispersion of the solution. Middle panel shows the CEL in a logarithmic scale, obtained averaging $\psi(t,z)$ in $z$ for each value of age. In this case, thin lines represent the interval containing 68\% of the distribution. Vertical dot-dashed lines show the times by which 10\%, 50\% and 95\% of the mass has been converted into stars. Bottom panel shows the normalized cumulative mass converted into stars. 
\label{f09}}
\end{figure}

\begin{figure}
\centering
\includegraphics[width=15cm,angle=0]{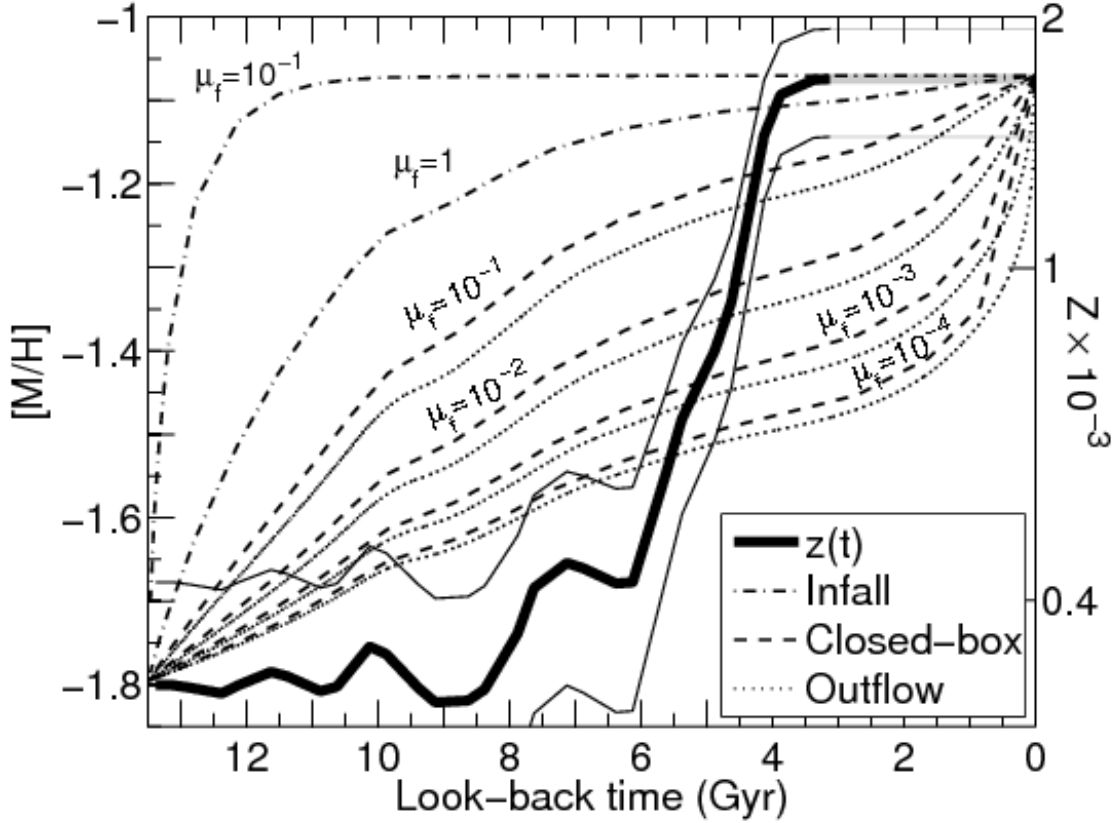}
\protect\caption[ ]{The CEL of Phoenix (thick, black line; thin black lines show the dispersion interval, see Figure \ref{f09}) compared with the results of several metallicity law models. Closed-box (dashed lines) for several final gas fractions $\mu_f$ are shown. Outflow models (dotted lines) with $\lambda=1$ are shown for the same $\mu_f$ used for the former. Infall models (dash-dotted lines) with $\alpha=1$ are also shown for $\mu_f=0.1$ and $\mu_f=1$. Only close-box and outflow models with $\mu_f=10^{-4}\sim 10^{-2}$ are compatible with the CEL of Phoenix. \citep*[See][for the definition of $\alpha$ and $\lambda$ parameters]{pei_col_sar1994}

\label{f10}}
\end{figure}

\notetoeditor{PLEASE, PUT FIGURES \ref{f08}, \ref{f09} AND \ref{f10} HERE}

For more clarity, Fig. \ref{f09} shows three different aspects of the SFH of Phoenix. In the upper panel the SFH as a function of time only $\psi(t)$ is shown. It is the same represented in the right-hand side, vertical panel of Fig. \ref{f08}. Thin lines correspond to the rms dispersion computed as explained above. The middle panel shows the chemical enrichment law (CEL) in a logarithmic scale. It is obtained from the averaged metallicity of the solution for each age. More precisely, this averaged metallicity is:

\begin{equation}
Z(t)=\frac{\int^{Z_{max}}_0 z'\psi(t,z')\,{\rm d} z'}{\int^{Z_{max}}_0 \psi(t,z')\,{\rm d} z'}
\end{equation}

\noindent where $Z_{max}=0.002$ is the maximum metallicity considered. The variable represented in Fig. \ref{f09} is the logarithm of $Z(t)$ in solar units, assuming $Z_\odot=0.0192$. Thin lines show the interval containing 68\% of the former integral. The mean metallicity for stars older than $\rm \sim 2~Gyr$ obtained from $Z(t)$ is $Z=0.0004\pm 0.0001$, which is compatible with the mean metallicity of the RGB stars obtained in \S\ref{secdistancia} using other method. Finally, bottom panel of Fig. \ref{f09} shows the normalized cumulative mass converted into stars obtained by simple integration of $\psi(t)$. 

We have obtained the SFH of Phoenix using models from the BaSTI stellar evolution library. However, a different set of stellar evolution models could produce a significantly different solution \citep{apa_hid2009, gal_zoc_apa2005}. To check the stability of our solution, we have calculated once more the SFH of Phoenix but using the stellar evolution models from Padua \citep{ber_etal1994} as input for IAC-star to compute the sCMD. The rest of input parameters are the same as those corresponding to the adopted solution, in particular 40\% of binaries and observational CMD offsets of $\Delta(V-I)=0.03$ and $\Delta I=0.075$. The solution based on the Padua library shows a faster SFR during the initial 2--3 Gyrs and is very similar from that epoch on. Differences are relatively small and do not affect the conclusions of this paper. Table \ref{tabsfh} gives a summary of the mean values derived from $\psi(t,z)$ for the BaSTI and Padua libraries. Column 1 identifies the stellar evolution library used. Column 2 gives the integral of $\psi(t,z)$ for the complete range of age and metallicity, that is, the total mass ever converted into stars. Columns 3 to 5 give the mean value for $\psi(t)$, metallicity, and age.

\notetoeditor{PLEASE, PUT TABLE \ref{tabsfh} HERE}

Our results (Figs. \ref{f08} and \ref{f09}) indicate that the Phoenix dwarf galaxy has had ongoing but gradually decreasing star formation over nearly a Hubble time. The highest level of star formation occurred from the formation of the galaxy and till 10.5 Gyr ago, when 50\% of the total star formation had already taken place. From that moment, star formation continues at a significant level until 6 Gyr ago (and additional 35\% of the stars are formed in this time interval), and at a very low level till the present time \citep[the youngest Phoenix stars are known to be $\simeq$ 100 Myr old;][]{mar_gal_apa1999}. These results are in good agreement with \citet{hol_smi_gri2000} and \citet{you_etal2007} who used the WFPC2 data for the inner field of the galaxy \citep[somewhat shallower dataset in the case of][]{hol_smi_gri2000}. Their intermediate-age star formation rate, however, seems to be somewhat higher, and in fact, 80\% of the stellar mass had formed 3 and X Gyr ago, respectively. The fact that their SFH is restricted to the inner field is however consistent with this trend, since we also find more intermediate-age star formation toward the center of the galaxy (See \S\ref{sfhradio}).

The CEL shown in Figure \ref{f09}, central panel, shows a trend of slowly increasing metallicity as a function of time till 8--6 Gyr ago, when Z increases steeply to the current value. From this CEL, we will try to extract some information about the physical conditions of the chemical enrichment in Phoenix. In Figure \ref{f10}, the Phoenix CEL is compared with several simple metallicity law models. It is apparent that none of the models adequately reproduce the Phoenix CEL. However, a few constraints may still be set. Infall models are the most clearly departing from the Phoenix CEL along the full history of the galaxy. We can therefore speculate that such process has likely not been relevant in the Phoenix history. A possible way to qualitatively explain the behavior of the Phoenix CEL during the first half of its history is by mass-loss of not mixed enriched material as it is ejected from dying stars. As for the second half, a fast enrichment is possible if fresh, unprocessed gas is very scarce and stars are born from gas heavily contaminated by stellar ejecta. The step down in the SFR at $\sim 6$ Gyr ago could be related with the same gas shortage. This qualitative scenario is compatible both with the fact that Phoenix has very little gas \citep{you_etal2007}, although it holds $\sim 100$ Gyr old stars.  However, this is only a qualitative approach and more detailed models must be computed to better account for the process.

\notetoeditor{PLEASE, PUT FIGURE \ref{f10} HERE}

\section{The SFH of Phoenix as a function of galactocentric radius}\label{sfhradio}

The study of the spatial distribution of the stellar populations provides fundamental information about the formation and evolution of dwarf galaxies. Low surface brightness structures play a significant role in this issue. Being fossil records of the early stages of the evolution of the galaxies, they can provide key information about the composition and structure of dwarf galaxies when they started to form stars. In particular they can reveal if a real halo formed only by old, low metallicity stars is present (see \S\ref{secintroduccion}). To this end, stellar age distribution must be accurately known in order to reach well-founded conclusions. This information is contained in the SFH $\psi(t,z)$. As we mentioned in \S\ref{secintroduccion}, there are evidences that Phoenix has a stellar population gradient. In this section we will study such a gradient by solving $\psi(t,z)$ as a function of the galactocentric radius. 

To derive the SFH as a function of radius, the observed stars were assigned to different regions of the galaxy. These regions were defined by fitting isophotes to ground based images of Phoenix \citep{mar_gal_apa1999} obtained with the LCO 100-inch telescope (Chile). To fit the isophotes, images observed in $I$ were combined and the resulting image was smoothed with a Gaussian filter of $\sigma=50$ pixels ($\sim 13''$). Ellipses were fitted to the isophotes with  ellipticity, position angle, and center as free parameters. A set of 6 ellipses of increasing semi-major axis were selected to divide the field in regions, each one including a statistically representative number of stars. 

\begin{figure}
\centering
\includegraphics[width=12cm,angle=0]{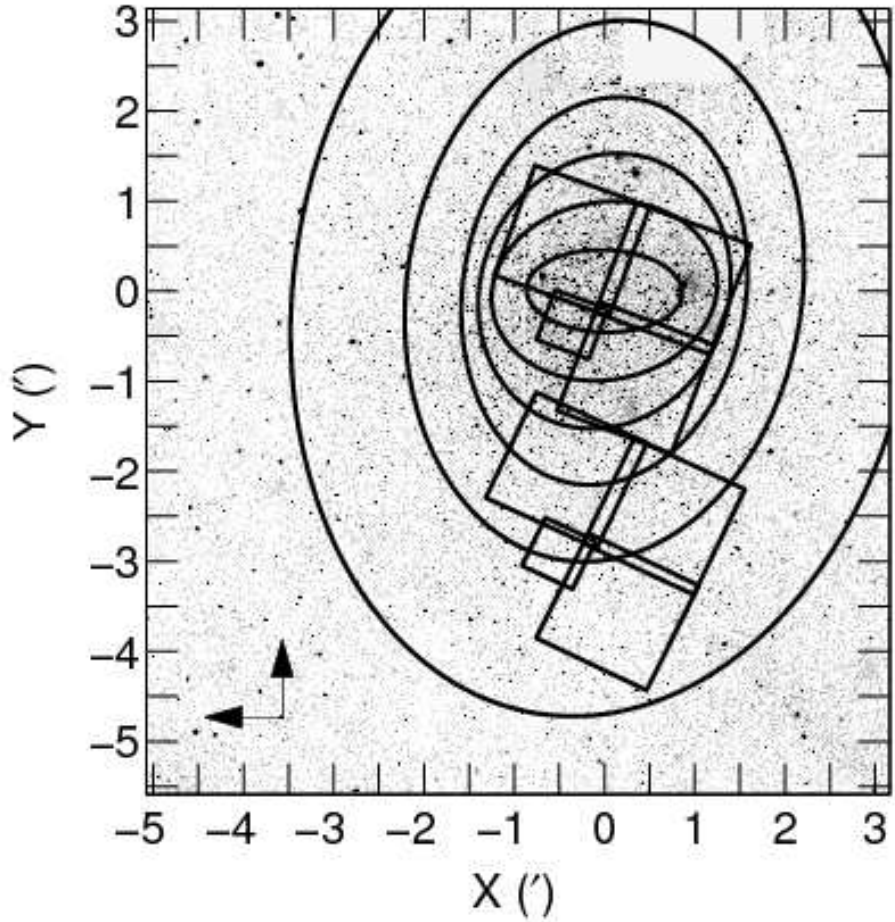}
\protect\caption[ ]{Ellipses selected to divide the galaxy field in regions for the study of stellar populations gradients on the image. The WFPC2 fields are also shown\label{f11}}
\end{figure}

\begin{figure}
\centering
\includegraphics[width=12cm,angle=0]{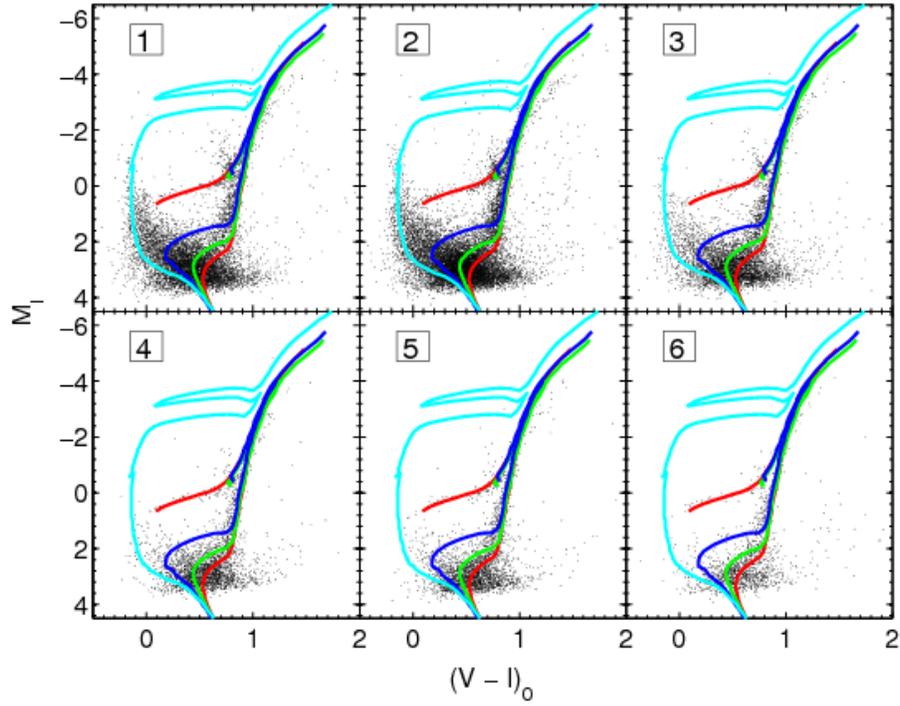}
\protect\caption[ ]{CMD of Phoenix for each ellipse shown in Fig. \ref{f11}. Isochrones of $\rm t=14~ Gyr, ~z=0.0001$ (red); $\rm t=8~ Gyr,~ z=0.004$ (green); $\rm t=5~ Gyr,~ z=0.004$ (blue); and $\rm t=0.2~ Gyr,~ z=0.0001$ (cyan) have been overplotted. Labels 1 to 6 refer to the elliptical regions shown in Fig. \ref{f11}, radius increasing from 1 to 6. 
\label{f12}}
\end{figure}

Figure \ref{f11} shows the selected ellipses overplotted on the ground based image of the galaxy. Figure \ref{f12} shows the six CMDs for the elliptical regions shown in Fig. \ref{f11}. Table \ref{elipsespar} shows the fitted parameters of the selected ellipses and the number of stars within the corresponding elliptical region. For each ellipse (\# 1 is the inner one), column 2 gives the semi-major axis in pc, column 3 gives the position angle, column 4 gives the ellipticity, and column 5 gives the number of stars between that ellipse and the former. Note the $\sim 90^\circ$ change in orientation between ellipses \#2 and \#4, which was already noted by \citet{mar_gal_apa1999} and that suggests two different galactic components. 
 
\notetoeditor{PLEASE, PUT FIGURE \ref{f11} HERE}
\notetoeditor{PLEASE, PUT FIGURE \ref{f12} HERE}
\notetoeditor{PLEASE, PUT TABLE \ref{elipsespar} HERE}

\begin{figure}
\centering
\includegraphics[width=15cm,angle=0]{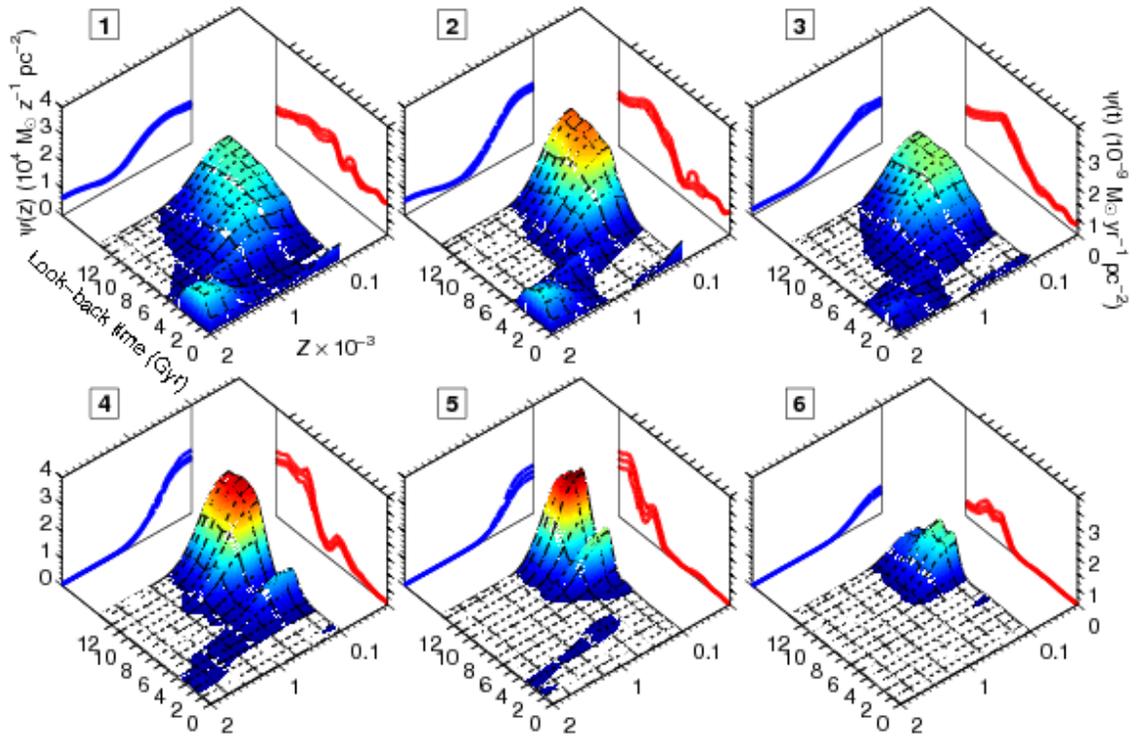}
\protect\caption{$\psi(t,z)$ for the elliptical regions shown if Fig. \ref{f11}. The radius increases from panel \#1 to \#6 (See Table \ref{elipsespar}). \label{f13}}
\end{figure}

\begin{figure}
\centering
\includegraphics[width=15cm,angle=0]{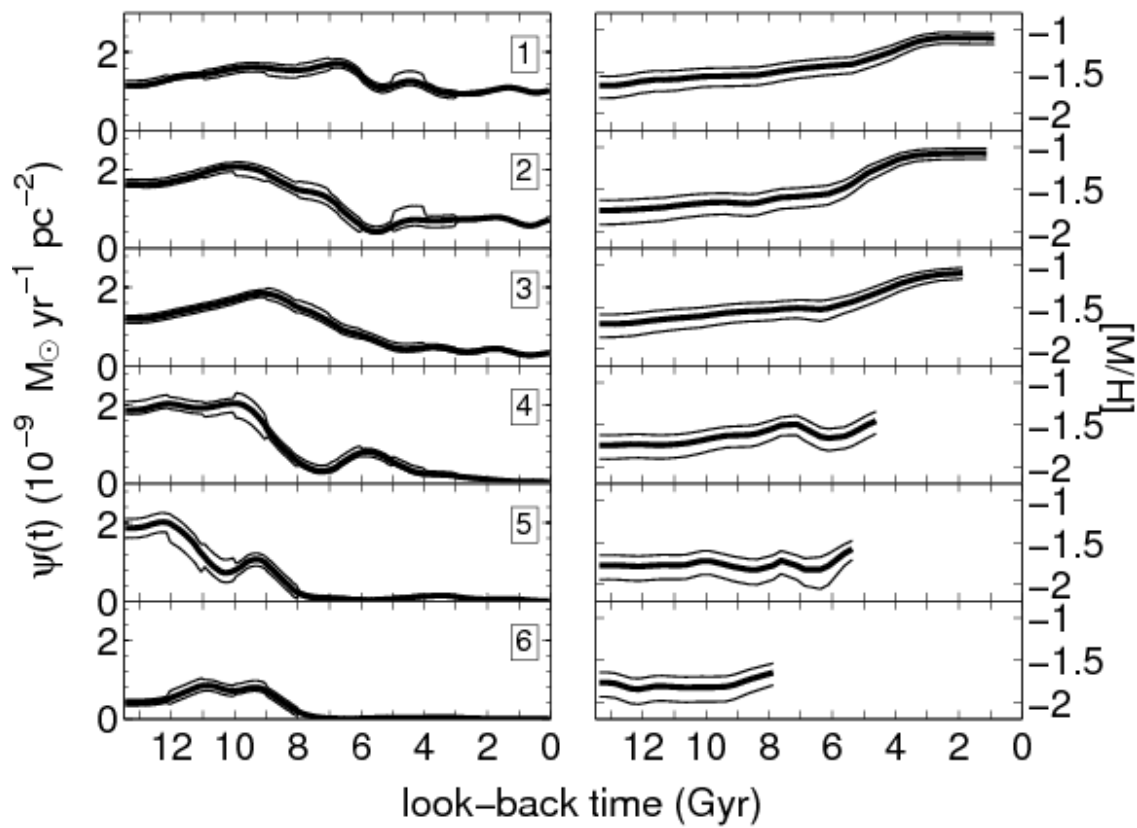}
\protect\caption[ ]{SFR (left panel) and CEL (right panel) for the six elliptical regions defined in Figure \ref{f11}. Regions are labeled as in Figures \ref{f12} and \ref{f13}. A boxcar average of width 1 Gyr and step of 0.1 Gyr has been used to smooth $\psi(t)$ and $z(t)$. In the CEL panel, metallicity law ends where no significant SFR is found.\label{f14}}
\end{figure}

The SFHs for CMDs shown in Fig. \ref{f12}, were solved as described in \S\ref{secsfh}. In particular, the observational effects were simulated on the sCMD according to the completeness tests for each region.  Solutions are shown in Fig. \ref{f13}. Figure \ref{f14} shows $\psi(t)$ (left panel) and the CEL (right panel) for each elliptical region. The CEL plots stop at ages for which no significant star formation is found. Regions \#1, \#2, and \#3 represent the main contribution to the global SFH. Regions \#4 to \#6 show mostly star formation at old and intermediate ages but the metallicities are similar to those of inner regions for the same age.

\notetoeditor{PLEASE, PUT FIGURE \ref{f13} HERE}
\notetoeditor{PLEASE, PUT FIGURE \ref{f14} HERE}

\begin{figure}
\centering
\includegraphics[width=12cm,angle=0]{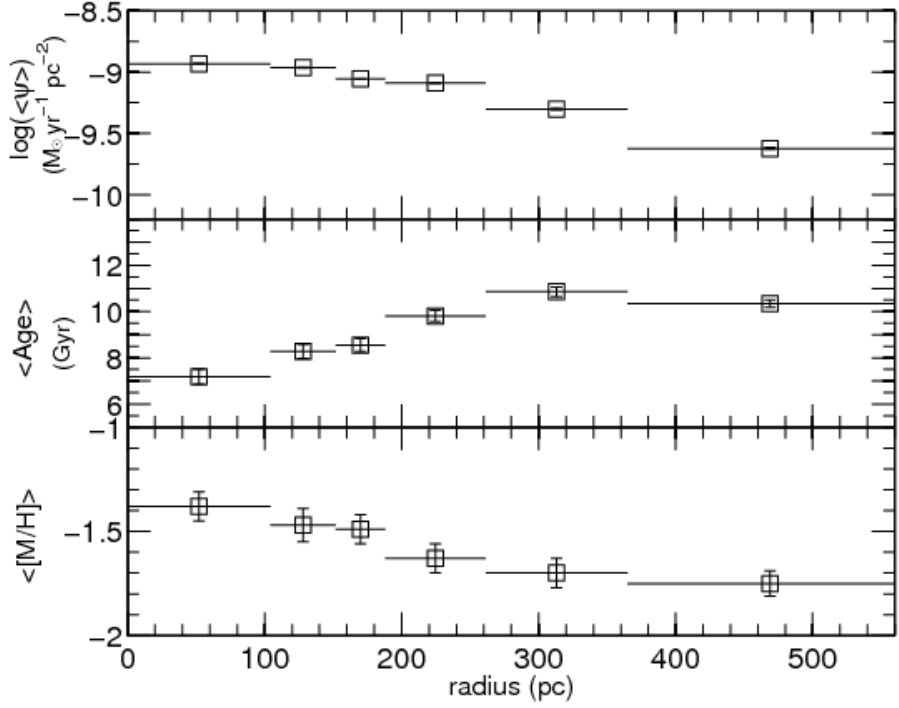}
\protect\caption[ ]{Upper, middle and bottom panels show respectively the mean $\psi(t)$, mean age and mean metallicity of the stars as a function of galactocentric radius. In all the cases, vertical bars are rms dispersions and horizontal bars show in the interval of age embraced by the associated value. 
\label{f15}}
\end{figure}

Figure \ref{f15} illustrates the stellar population gradient in Phoenix. Upper, middle, and lower panels show the mean values of the SFR ($<\psi>$), age, and metallicity ($<[M/H]>$) as a function of the galactocentric distance. Although a step down is observed in $<\psi>$ in the outermost region, its behaviour is compatible with a single exponential decrease, without any significant change of slope. $<[M/H]>$ goes slightly down in the inner 200 pc and remains approximately constant from that distance on. 

Table \ref{meanreg} gives a summary of the main properties of the SFH of Phoenix as a function of galactocentric distance. The limits of the semi-major axis (in pc) for each elliptical region is given in column 1; columns 2 to 5 give the values of $<\psi>$, $<[M/H]>$, and $<age>$, respectively.

\begin{figure}
\centering
\includegraphics[width=12cm,angle=0]{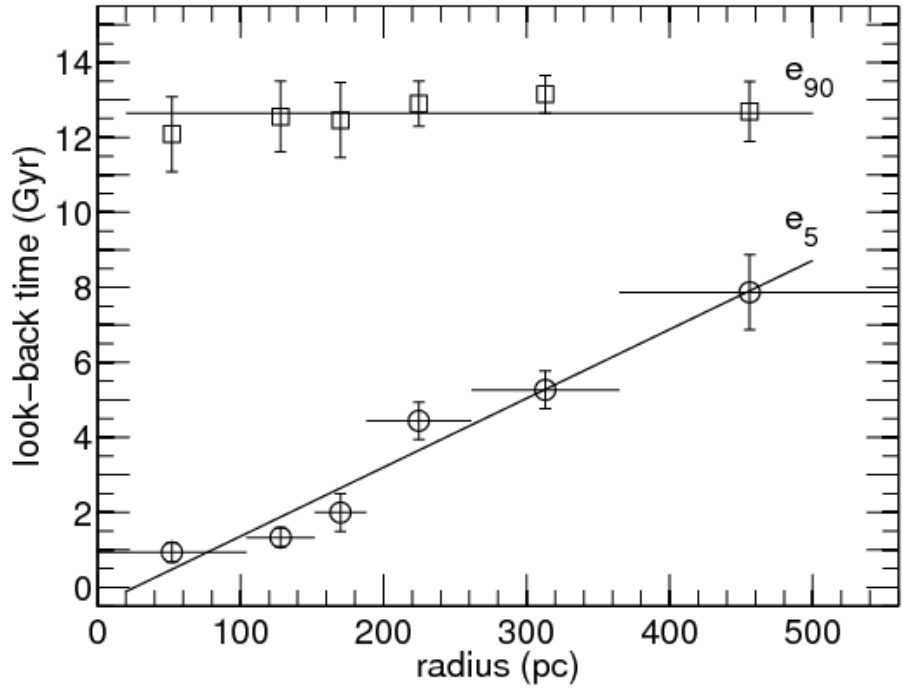}
\protect\caption[ ]{The 90-percentile ($e_{90}$) (open squares) and 5-percentile ($e_5$) (open circles) of the star age distribution. In both cases, vertical bars are rms dispersions and horizontal bars show the interval of age embraced by the associated value. 
\label{f16}}
\end{figure}

\notetoeditor{PLEASE, PUT FIGURE \ref{f15} \ref{f16} \ref{f17} \ref{f18} HERE}
\notetoeditor{PLEASE, PUT TABLE \ref{meanreg} HERE}

Figure \ref{f16} shows the ages corresponding to the 90th-percentile ($e_{90}$) and the 5th-percentile ($e_5$) of the age distribution. $e_{90}$ is the age such that 90\% of the star formation has happened more recently than it, and similarly $e_5$. In other words, $e_{90}$ and $e_5$ are indicative of the age at which the star formation start and finishes, respectively. The flat distribution of $e_{90}$ shows that the oldest population has almost the same age all over the galaxy. The steep behaviour of $e_5$ shows that the younger the stars, the more concentrated they are. But, in relation to the existence or not of a real old halo in Phoenix, the most interesting characteristic of the $e_5$ behaviour is that it shows a smooth variation with no sharp step. 

In summary, all together, the smooth gradients or flatness of $<\psi>$, $<[M/H]>$, $<age>$, $e_{90}$, and $e_5$ are difficult to reconcile with the presence of an halo in Phoenix, as a break in the stellar population gradients would be expected in such a case \citep*{tik_gal_dro2005}. The presence of intermediate age stars in the external regions neither supports the presence of an old halo in the galaxy. The change in the orientation of the ellipses observed between regions \#2 and \#4 does not correlate with any sharp change in the slope of any of the mean and percentile values obtained here. The orientation of the inner ellipses could be then related to the distribution of the youngest stars in the last star formation event. Therefore, we can state that at a distance of about 500 pc from the center, no traces of an old halo are observed in Phoenix.

\section{The evolution of the galaxy structure}\label{evogal}

The behaviour of $e_5$ (Fig. \ref{f16}) shows that the age of youngest stars decreases as we move to inner regions in the galaxy. This could be produced if stars form in an inner region and then migrate to outer ones. Alternatively, if stars are formed {\it in situ}, it is compatible with a shrinking star forming region centered in the galactic center and initially extending out to the galaxy outskirts. Contraction of this region would be the simple consequence of pressure support reduction as gas supply diminishes. As we have already mentioned, \citet{stin_etal2009} have shown on the basis of theoretical models that the latter would be the most efficient mechanism at work in dwarf galaxies.

\begin{figure}
\centering
\includegraphics[width=12cm,angle=0]{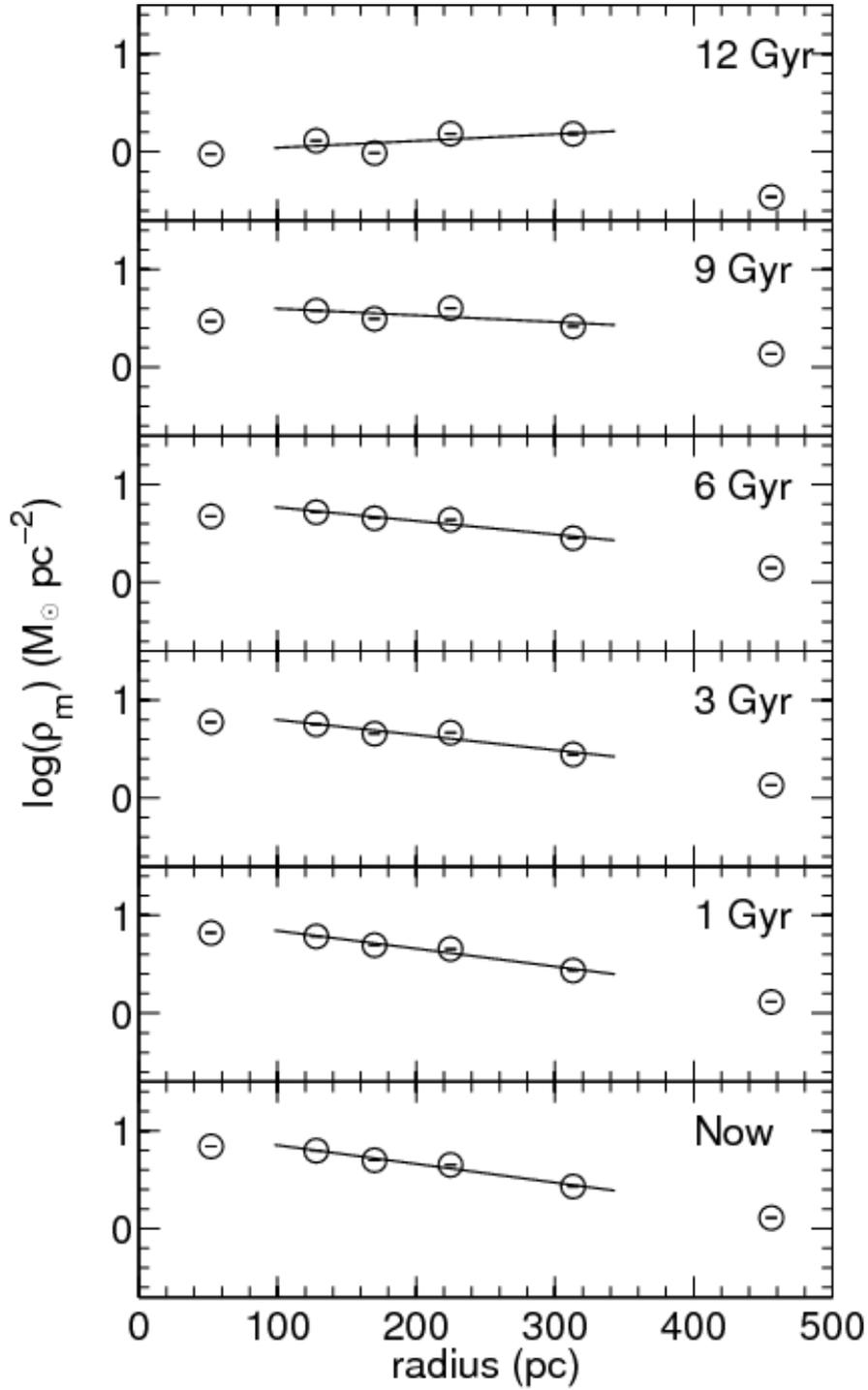}
\protect\caption[ ]{Distributions of surface mass density in alive stars as a function of galactocentric radius for several look-back times in Phoenix. Lines show fits to data for ellipses 2 to 5.  
\label{f17}}
\end{figure}

Our results for Phoenix allow us to study the way in which the galaxy evolves and, in the case that stars are formed {\it in situ}, to compute the rate at which the star forming region shrinks. The morphological evolution of the galaxy can be described in terms of what would be seen if it would be observed at different look-back times or red-shifts. To this purpose, the SFRs shown in Fig. \ref {f14} have been given as input to IAC-star to compute the corresponding synthetic populations as would be observed at different times. The surface mass density in alive stars as a function of radius and time, $\rho_m(r,t)$, has been derived using time steps of 1 Gyr. Results are shown in Figure \ref{f17} for a sample of six values of look-back time. The most noticeable results from this figure are the reverse slope for the early epoch and the break at about 350 to 450 pc, also for the early epoch. A small depletion is also visible in the innermost ellipse at all epochs, which has been found in other dwarf galaxies (see e.g. \citealt{apa_tik2000, apa_tik_kar2000, apa_etal2001}). Exponential fits to data corresponding to ellipses 2 to 5 are also shown. Innermost and outermost ellipses have been excluded due to the mentioned departures from the intermediate ellipses trend. 

\begin{figure}
\centering
\includegraphics[width=12cm,angle=0]{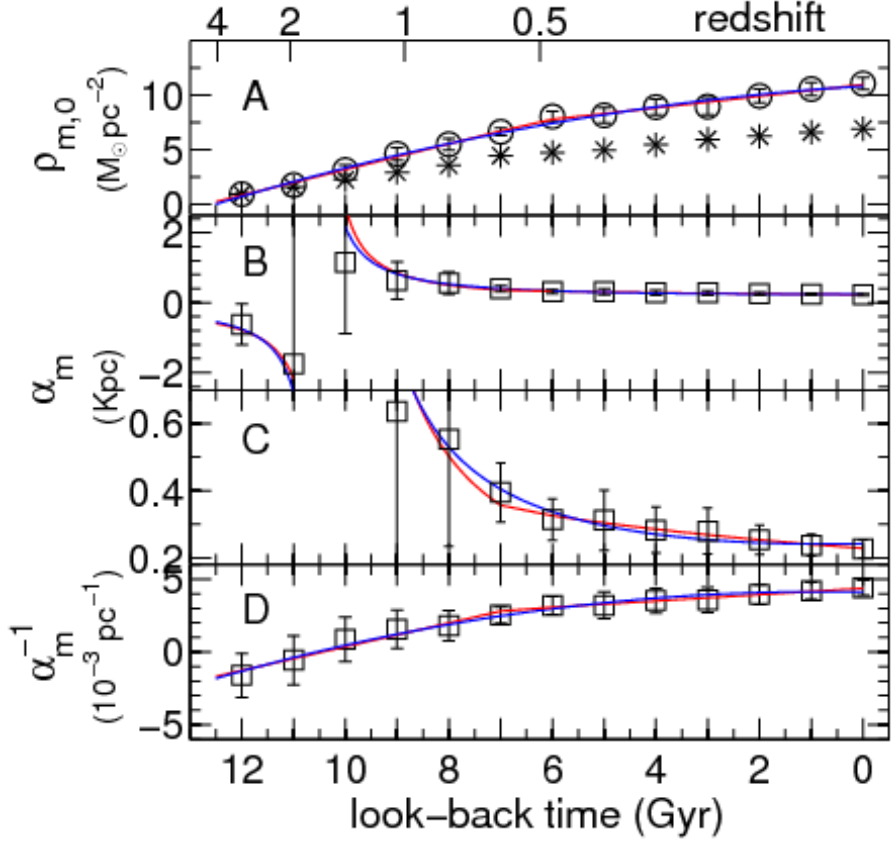}
\protect\caption[ ]{Central surface density and scale length of the distribution of surface mass density in alive stars as a function of look-back time in Phoenix. Panel (A) shows the central density; open circles represent the values obtained from the fit to the alive star mass distribution, while asterisks provide the average values directly measured in the inner ellipse (see Fig. \ref{f11}). Red line shows two linear fits performed to the values corresponding to look-back times larger and smaller than 6.5 Gyr, respectively. Blue lines represent a quadratic fit. Panel (B) shows the scale length. Panel (C) shows the same with a different vertical scale, for more detail. For illustrative purposes, Panel (D) shows the inverse of the scale length, to emphasize the fact that the slope of the alive star mass distribution changes sign at about 10.6 Gyr ago. In panels (B) to (D), red lines show two linear fits to $\alpha_m^{-1}$ performed to the values corresponding to look-back times larger and smaller than 6.5 Gyr, respectively, while blue lines represent a quadratic fit, also to $\alpha_m^{-1}$.
\label{f18}}
\end{figure}

The scale length, $\alpha_m(t)$, and central density, $\rho_{m,0}(t)$, both as a function of time, have been computed from fits similar to the ones shown in Figure \ref{f17} but using all the available time snapshots (from 0 to 12 Gyr with 1 Gyr time step). Results are plotted in Figure \ref{f18} which, for more clarity, shows $\alpha_m(t)$ in two different scales as well as $\alpha^{-1}_m(t)$, i.e. the slope of the $\rm \ln \rho_m(r,t) vs. {\rm r}$ fit. A red-shift scale is also given according to the cosmological parameters for a flat Einstein-de Sitter universe and the 3-years WMAP data \citep{spergel_etal2007}.

There are several interesting things to note from this figure. The most striking one, already seen in Figure \ref{f17} is that the slope of the mass density distribution, $\alpha^{-1}_m(t)$, was negative in the early epoch and changed sign at about 10.6 Gyr ago. In other words, mass density in stars was higher in outer regions than in inner ones during the early phase of the galaxy evolution. Besides that, the scale length of the alive stars component mass has steadily and smoothly decreased since 11 Gyr ago while the central density has increased since the earliest epoch. Quadratic fits can be done to $\alpha^{-1}_m(t)$ and $\rho_{m,0}(r,t)$, but it is remarkable that linear fits with a break somewhere between 6 and 7 Gyr ago also provide a good representation of data. This epoch is coincident with the time at which chemical enrichment departs from a closed box (see Section \ref{secsfh}). We have pointed to the possibility that fresh gas consumption would have been the cause of the latter. If shrinking is produced by the reduction of gas supply, it is not surprising that fresh gas consumption would produce some change in the shrinking rate. However, theoretical models are necessary to ascertain if this is the case. 

In summary, if stars form {\it in situ} in Phoenix, Figures \ref{f17} and \ref{f18} indicate that the onset of star formation took place all over the galaxy, up to a distance of about 400 pc from the center at least, and with a relatively larger intensity in intermediate to outer regions, rather that in the innermost ones. After a few Gyr, star formation in inner regions started to dominate. Besides that, the star formation has become more and more concentrated, slowly shrinking down as time goes on. This scenario shows good agreement with model results by \citet{stin_etal2009} and seems to rule out the possibility that the galaxy holds a true old halo or, at least, that it could be significant at 500 pc from the center, equivalent to about 2.5 scale-lengths. 

\section{Summary and conclusions}\label{concl}

Photometry of resolved stars for the Phoenix dwarf galaxy using WFPC2@HST with the F555W and F814W filters  have been presented. Color-magnitude diagrams, distance, mean metallicity, and the SFH, both global and as a function of radius has been obtained. This information allowed us to study the main features of the galaxy and the spatial distribution of its stellar populations. Results can be summarized as follows:

\paragraph{(1)}
Using the TRGB, a distance modulus of $(m - M)_0 = 23.09\pm 0.10$, corresponding to $d=415\pm 19$ kpc has been obtained. A mean metallicity $[Fe/H]=-1.87\pm 0.06$ dex, ($Z=(2.6\pm 0.3)\times 10^{-4}$) for the RGB stars has been also derived.

\paragraph{(2)}
Using IAC-pop, the star formation rate as a function of time and metallicity, $\psi(t,z)$, has been obtained, globally and as a function of radius. Phoenix has stars of all ages with a metallicity range of $Z=0.0001-0.002$. Intermediate age and old stars can be found at all galactocentric distances.

\paragraph{(3)}
Phoenix has had ongoing but gradually decreasing star formation over nearly a Hubble time. Star formation started in Phoenix at a very early epoch: 10\% of the stars ever born in Phoenix were already formed about 13 Gyr ago.  50\% of the total star formation had already taken place by 10.5 Gyr ago. Then, star formation continued at a significant level until 6 Gyr ago, forming an additional 35\% of the stars, and at a very low level till around 100 Myr ago. 

\paragraph{(4)}
The distribution of mass density in alive stars and its evolution with time has been studied. 
If the stars we observe today were formed {\it in situ} in Phoenix, the onset of the star formation took place all over the galaxy, to a distance of at least 400 pc from the center. A relatively larger rate is observed in intermediate to outer regions in the earliest star formation episode, indicating that the star formation onset occurred preferentially in those external regions.

\paragraph{(5)}
After a few Gyrs, the star formation in inner regions started to dominate, although it continued at a relatively high rate also in intermediate and outer regions for several Gyrs. At about 7 to 8 Gyrs ago, star formation stopped in the outermost regions (about 0.5 kpc from the center). From that point on, it was progressively stopping closer to the central regions as time went on, until the present, in which recent residual star formation activity is observed in the innermost region. 
As a result of this process, the scale-length of the stellar mass density distribution decreases with time. 

\paragraph{(6)}
The former sketch is compatible with a scenario in which the star formation region envelope slowly shrinks as time goes on, possibly as the natural result of pressure support reduction as gas supply diminishes. This results in a) old and intermediate-age stars present at all galactocentric distances but young stars only in the inner regions, b) a positive galactocentric gradient of age of the youngest population and a positive small gradient in the mean stellar age. All these results are in agreement with our observations. 

\paragraph{(7)}
The chemical enrichment law has been obtained from $\psi(t,z)$. Chemical enrichment was very slow until about 6 Gyr ago, when metallicity increased abruptly. The rough coincidence with the time at which the star forming region envelope start to shrink allows us to speculate that, from that time on, the fraction of fresh, unprocessed gas involved in subsequent star formation dramatically decreased. Detailed theoretical models are necessary to ascertain the plausibility of this suggestion. 

\paragraph{(8)}
No traces of a true, old halo are apparent in Phoenix either in its stellar age distribution or in the stellar mass density distribution, at least out to 0.5 kpc (about 2.5 scale-lengths) from the center. The time evolution of the mass density distribution does not show traces of such structure either.

\acknowledgments

We are very grateful to E. Skillman, P.~B. Stetson, and K. McQuinn for their suggestions to improve this paper. The authors are funded by the IAC (grant P3/94) and by the Science and Technology Ministry of the Kingdom of Spain (grants AYA2004-3E4104 and AYA2007-3E3507).

\clearpage

\clearpage

\clearpage

\clearpage

\clearpage

\clearpage

\clearpage

\clearpage

\clearpage

\clearpage

\clearpage

\clearpage

\clearpage

\clearpage

\clearpage

\clearpage

\clearpage

\clearpage

\clearpage
\begin{deluxetable}{lcc}
\tabletypesize{\scriptsize}
\tablecaption{Main properties of Phoenix dwarf galaxy.\label{phoenixprop}}
\tablewidth{0pt}
\tablehead{
\colhead{Property} &\colhead{Value} &\colhead{Refs.}}
\startdata          
Galactic coordinates ($l,b$)   & 272\fdg19, -68\fdg95 &5\\
Equatorial coordinates ($\alpha_{2000}$,$\delta_{2000}$)    & $\rm 01^h51^m03\fs3$, $-44\arcdeg 27\arcmin 11\arcsec$  &5\\
Distance modulus $(m-M)_0$    &$23.1\pm 0.1$, $23.0\pm 0.1$, $23.21\pm 0.08$, $23.11$, $23.09\pm 0.10$ &6,4,2,3,7\\
Reddening $A_V$,$A_I$  &0.062, 0.036 &1\\
Absolute Magnitude $M^0_V$ &$-9.8\pm 0.4$ &4\\
Total mass ($10^6~M_\odot$) &$4.4\pm 0.3$ &7\\
$\rm [Fe/H]$ &$-1.81\pm 0.10$ &2,3\\
Average star formation rate ($10^{-9}M_\odot yr^{-1} pc^{-2}$)  &$0.99\pm 0.08$ &7\\
\enddata
\tablerefs{(1) Cardelli, Clayton, \& Mathis 1989; (2) Held, Saviane, \& Momany 1999; (3) Holtzman, Smith, \& Grillmair 2000; (4) Mart\'{\i}nez-Delgado, Gallart, \& Aparicio 1999; (5) Schuster \& West 1976; (6) van de Rydt, Demers, \& Kunkel 1991; (7) This work.}
\end{deluxetable}
\clearpage

\begin{deluxetable}{cccccccc}
\tabletypesize{\scriptsize}
\tablecaption{Observation Log.\label{orbphoenix}}
\tablewidth{0pt}
\tablehead{
\colhead{Visit} & \colhead{Field} & \colhead{Cycle} & \colhead{Proposal ID} & \colhead{Number of orbits} & \colhead{Filter}& \colhead{Number of images} & \colhead{Exp. time (s)}}
\startdata
2 &inner &6 &6798 &2   &F555W  &~6  &~4900\\
2 &inner &6 &6798 &2   &F814W  &~6  &~4900\\
1 &inner &9 &8796 &3   &F555W  &~6  &~7200\\
2 &inner &9 &8796 &4   &F814W  &~8  &~9600\\
3 &outer &9 &8796 &5   &F555W  &11  &11900\\
4 &outer &9 &8796 &6   &F814W  &13  &14300\\
\enddata
\end{deluxetable}
\clearpage

\begin{deluxetable}{cccccc}
\tabletypesize{\scriptsize}
\tablecaption{Photometric calibration terms.\label{calphoenix}}
\tablewidth{0pt}
\tablehead{
\colhead{\# CCD}  &\colhead{$Z$} & \colhead{$C$} & \colhead{$K\times 10^{-4}$} &\colhead{$\sigma$}}
\startdata
PC (V)   &21.623 &0      &1.27 &0.012\\
PC (I)   &20.829 &-0.053 &0.94 &0.004\\
WFC2 (V) &21.740 &-0.032 &0    &0.003\\
WFC2 (I) &20.888 &-0.023 &0    &0.004\\
WFC3 (V) &21.796 &-0.063 &0    &0.003\\
WFC3 (I) &20.896 &-0.033 &0    &0.004\\
WFC4 (V) &21.657 & 0     &2.19 &0.003\\
WFC4 (I) &20.840 &-0.030 &1.37 &0.004\\
\enddata
\end{deluxetable}
\clearpage

\begin{deluxetable}{cccccc}
\tabletypesize{\scriptsize}
\tablecaption{Mean and integrated values derived from $\psi(t,z)$.\label{tabsfh}}
\tablewidth{0pt}
\tablehead{
\colhead{SEL} &\colhead{Total Mass } &\colhead{$<\psi>$} &\colhead{$\rm <[M/H]>$} &\colhead{$<age>$}\\
\colhead{} &\colhead{($10^6~ M_\sun$)}&\colhead{($10^{-9} M_\sun yr^{-1} pc^{-2}$)}  &\colhead{}  &\colhead{(Gyr)}}
\startdata
BaSTI &$4.4\pm 0.3$ &$0.99\pm 0.08$ &$-1.7\pm 0.2$ &$~\,9.4\pm 1.1$\\
Padua &$4.8\pm 0.4$ &$1.04\pm 0.07$ &$-1.7\pm 0.1$ &$10.0\pm 1.2$\\
\enddata
\end{deluxetable}
\clearpage

\begin{deluxetable}{ccccc}
\tabletypesize{\scriptsize}
\tablecaption{Fitted parameters for the ellipses in Fig. \ref{f11}.\label{elipsespar}}
\tablewidth{0pt}
\tablehead{
\colhead{ellipse \#} & \colhead{sma (pc)} & \colhead{$\theta\rm (^\circ)$} & \colhead{$\epsilon$} & \colhead{$N_\star$}}
\startdata
1 &104.2  &-92 &0.5 &8640\\
2 &151.8 &-80 &0.2 &12235\\
3 &199.0 &-24 &0.1 &5578\\
4 &261.4 &-9  &0.3 &3011\\
5 &364.8 &-9  &0.3 &3013\\
6 &573.2 &-9  &0.3 &1585\\
\enddata
\end{deluxetable}

\clearpage

\begin{deluxetable}{ccccc}
\tabletypesize{\scriptsize}
\tablecaption{Mean values as a function of radius.\label{meanreg}}
\tablewidth{0pt}
\tablehead{
\colhead{sma} &\colhead{$<\psi>$} &\colhead{$\rm <[M/H]>$} & \colhead{$<age>$}\\
\colhead{(pc)} &\colhead{($10^{-9} M_\sun yr^{-1} pc^{-2}$)} &\colhead{} &\colhead{(Gyr)}}
\startdata
~~~0 ~ -- 104.2 &$1.61\pm 0.09$     &$-1.57\pm 0.13$  &$~8.2\pm 1.3$\\
104.2 -- 151.8  &$1.42\pm 0.09$     &$-1.62\pm 0.14$  &$~8.8\pm 1.2$\\
151.8 -- 188.0  &$1.26\pm 0.09$     &$-1.66\pm 0.13$  &$~9.6\pm 1.1$\\
188.0 -- 261.4  &$1.08\pm 0.17$     &$-1.74\pm 0.11$  &$10.7\pm 0.9$\\
261.4 -- 364.8  &$0.64\pm 0.06$     &$-1.74\pm 0.13$  &$11.2\pm 0.7$\\
364.8 -- 573.2  &$0.30\pm 0.04$     &$-1.80\pm 0.08$  &$10.8\pm 0.7$\\
\enddata
\end{deluxetable}


\begin{thebibliography}{}


\bibitem[Aparicio et al.(2001)]{apa_etal2001}
Aparicio, A., Carrera, R, \& Mart{\'\i}nez-Delgado, D.\ 2001, \aj, 122, 2524

\bibitem[Aparicio et al.(1997)]{apa_etal1997}
Aparicio, A., Dalcanton, J.~J., Gallart, C., \& Mart{\'\i}nez-Delgado, D.\ 1997, \aj, 114, 1447

\bibitem[Aparicio \& Gallart(1995)]{apa_gal1995}
Aparicio, A., \& Gallart, C.\ 1995, \aj, 110, 2105

\bibitem[Aparicio \& Gallart(2004)]{apa_gal2004}
Aparicio, A., \& Gallart, C.\ 2004, \aj, 128, 1465

\bibitem[Aparicio \& Hidalgo(2009)]{apa_hid2009}
Aparicio, A., \& Hidalgo, S.~L.\ 2009, \aj, accepted.

\bibitem[Aparicio \& Tikhonov(2000)]{apa_tik2000}
Aparicio, A., \& Tikhonov, N.\ 2000, \aj, 119, 2183

\bibitem[Aparicio et al.(2000)Aparicio, Tikhonov, \& Karachentsev]{apa_tik_kar2000}
Aparicio, A., Tikhonov, N., \& Karachentsev, I.\ 2000, \aj, 119, 177

\bibitem[Bagget et al.(2002)]{bag_etal2002}
Baggett, S., et al. 2002, in HST WFPC2 Data Handbook, v. 4.0, ed. B. Mobasher, Baltimore, STScI

\bibitem[Battinelli \& Demers(2000)]{bat_dem2000} Battinelli, P., \& Demers, S.\ 2000, \aj, 120, 1801 

\bibitem[Battinelli \& Demers(2006)]{bat_dem2006} Battinelli, P., \& Demers, S.\ 2006, \aap, 447, 473 

\bibitem[Battinelli et al.(2007)Battinelli, Demers, \& Artigau]{bat_dem_art2007} Battinelli, P., 
Demers, S., \& Artigau, {\'E}.\ 2007, \aap, 466, 875 

\bibitem[Battinelli et al.(2006)Battinelli, Demers, \& Kunkel]{bat_dem_kun2006} 
Battinelli, P., Demers, S., \& Kunkel, W.~E.\ 2006, \aap, 451, 99 

\bibitem[Bellazzini et al.(2004)]{bel_etal2004} Bellazzini, M., Ferraro, F.~R., Sollima, A., Pancino, E., \& Origlia, L.\ 2004, \aap, 424, 199

\bibitem[Bertelli et al.(1994)]{ber_etal1994}
Bertelli, G., Bressan, A., Chiosi, C., Fagotto, F., \& Nasi, E.\ 1994, \aaps, 106, 275

\bibitem[Binggeli et al.(1990)Binggeli, Tarenghi, \& Sandage]{bin_tar_san1990}
Binggeli, B., Tarenghi, M., \& Sandage, A.\ 1990, \aap, 228, 42

\bibitem[Canterna \& Flower(1977)]{can_flo1977}
Canterna, R., \& Flower, P.~J.\ 1977 \apjl,  212, L57

\bibitem[Cardelli et al.(1989)Cardelli, Clayton, \& Mathis]{car_cla_mat1989} Cardelli, J.~A., 
Clayton, G.~C., \& Mathis, J.~S.\ 1989, \apj, 345, 245 

\bibitem[Carignan et al.(1991)Carignan, Demers, \& Cote]{car_dem_cot1991}
Carignan, C., Demers, S., \& Cote, S.\ 1991, \apjl, 381, L13

\bibitem[Castelli \& Kurucz(2004)]{cas_kur2004}
Castelli, F., \& Kurucz, R.~L.\ 2004, {\it New Grids of ATLAS9 Model Atmospheres} IAU Symposium 210, A20

\bibitem[Charbonneau(1995)]{cha1995} Charbonneau, P.\ 1995, 
\apjs, 101, 309

\bibitem[Dalcanton \& Bernstein(2002)Dalcanton \& Bernstein]{dal_ber2002} Dalcanton,
J.~J., \& Bernstein, R.~A.\ 2002, \aj, 124, 1328

\bibitem[Dekel \& Silk(1986)]{dek_sil1986}
Dekel, A., \& Silk, J.\ 1986, \apj, 303, 39

\bibitem[Demers \& Battinelli(2002)]{dem_bat2002} Demers, S., \& 
Battinelli, P.\ 2002, \aj, 123, 238 

\bibitem[Demers et al.(2006)Demers, Battinelli, \& Artigau]{dem_bat_art2006} 
Demers, S., Battinelli, P., \& Artigau, E.\ 2006, \aap, 456, 905 

\bibitem[Dolphin(1997)]{dol1997} Dolphin, A.\ 1997, New 
Astronomy, 2, 397 

\bibitem[Dolphin(2002)]{dol2002} Dolphin, A.~E.\ 2002, \mnras, 
332, 91 

\bibitem[Gallart et al.(2004a)]{gal_etal2004a} Gallart, C., Aparicio, 
A., Freedman, W.~L., Madore, B.~F., Mart{\'{\i}}nez-Delgado, D., \& 
Stetson, P.~B.\ 2004, \aj, 127, 1486 

\bibitem[Gallart et al.(2004b)]{gal_etal2004b} Gallart, C., Stetson, 
P.~B., Hardy, E., Pont, F., \& Zinn, R.\ 2004, \apjl, 614, L109 

\bibitem[Gallart et al.(2001)]{gal_etal2001} Gallart, C., 
Mart{\'{\i}}nez-Delgado, D., G{\'o}mez-Flechoso, M.~A., \& Mateo, M.\ 2001, 
\aj, 121, 2572 

\bibitem[Gallart et al.(2005)Gallart, Zoccali, \& Aparicio]{gal_zoc_apa2005} Gallart, C., Zoccali, 
M., \& Aparicio, A.\ 2005, \araa, 43, 387 

\bibitem[Girardi et al.(2000)]{gir_etal2000} Girardi, L., Bressan, A., Bertelli, G., \& Chiosi, C.\ 2000, \aaps, 141, 371 

\bibitem[Harris(1991)]{har1991}
Harris, W.~E.\ 1991, \araa, 29, 543

\bibitem[Harris \& Zaritsky(2001)]{har_zar2001} Harris, J., \& Zaritsky, D.\ 2001, \apjs, 136, 25 

\bibitem[Held(2002)]{hel2002} Held, E.~V.\ 2002, Scientific 
Drivers for ESO Future VLT/VLTI Instrumentation, 178 

\bibitem[Held et al.(1999)Held, Saviane, \& Momany]{hel_sav_mom1999}
Held, E.~V., Saviane, I., \& Momany, Y.\ 1999, \aap, 345, 747

\bibitem[Hidalgo et al.(2003)Hidalgo, Mar{\'{\i}}n-Franch, \& Aparicio]{hid_mar_apa2003}
Hidalgo, S.~L., Mar{\'{\i}}n-Franch, A., \& Aparicio, A.\ 2003, \aj, 125, 1247

\bibitem[Hill et al.(1998)]{hil_etal1998}
Hill, R.~J., Ferrarese, L., Stetson, P.~B., Saha, A., Freedman, W.~L., Graham, J.~A., Hoessel, J.~G., Han, M., Huchra, J., Hughes, S.~M., Illingworth, G.~D., Kelson, D., Kennicutt, R.~C., Bresolin, F., Harding, P., Turner, A., Madore, B.~F., Sakai, S., Silbermann, N.~A., Mould, J.~R. \& Phelps, R.,\ 1998, \apj, 496, 648

\bibitem[Hodge(1989)]{hod1989} Hodge, P.\ 1989, \araa, 27, 139 

\bibitem[Holtzman et al.(2000)Holtzman, Smith, \& Grillmair]{hol_smi_gri2000} Holtzman, J.~A., 
Smith, G.~H., \& Grillmair, C.\ 2000, \aj, 120, 3060 

\bibitem[Irwin \& Tolstoy(2002)]{irw_tol2002} Irwin, M., \& Tolstoy, E.\ 2002, \mnras, 336, 643 

\bibitem[Kroupa et al.(1993)Kroupa, Tout, \& Gilmore]{kro_tou_gil1993}
Kroupa, P., Tout, C.~A., \& Gilmore, G.\ 1993, \mnras, 262, 545

\bibitem[Komiyama et al.(2003)Komiyama et al.]{kom_etal2003}
Komiyama, Y, Okamura, S., Yagi, M., Furusawa, H., Doi, M. et al.\ 2003, \apj, 590, L17


\bibitem[Leisy et al.(2005)]{lei_etal2005} 
Leisy, P., Corradi, R.~L.~M., Magrini, L., Greimel, R., Mampaso, A., \& Dennefeld, M.\ 2005, \aap, 436, 437 


\bibitem[Letarte et al.(2002)]{let_etal2002} Letarte, B., Demers, 
S., Battinelli, P., \& Kunkel, W.~E.\ 2002, \aj, 123, 832 

\bibitem[Madore \& Freedman(1995)]{mad_fre1995}
Madore, B.~F., \& Freedman, W.~L.\ 1995, \aj, 109, 1645

\bibitem[Magrini et al.(2003)]{mag_etal2003} Magrini, L., et al.\ 2003, \aap, 407, 51 

\bibitem[Mart{\'{\i}}nez-Delgado et al.(1999)Mart{\'{\i}}nez-Delgado, Gallart, \& Aparicio]{mar_gal_apa1999} 
Mart{\'{\i}}nez-Delgado, D., Gallart, C., \& Aparicio, A.\ 1999, \aj, 118, 862 

\bibitem[Mayer et al.(2001)]{may_etal2001} Mayer, L., Governato, F., 
Colpi, M., Moore, B., Quinn, T., Wadsley, J., Stadel, J., \& Lake, G.\ 
2001, \apj, 559, 754 

\bibitem[Mayer et al.(2006)]{may_etal2006} Mayer, L., Mastropietro, 
C., Wadsley, J., Stadel, J., \& Moore, B.\ 2006, \mnras, 369, 1021 

\bibitem[Menzies et al.(2008)]{men_etal2008} Menzies, J., Feast, M., 
Whitelock, P., Olivier, E., Matsunaga, N., \& da Costa, G.\ 2008, \mnras, 385, 1045 

\bibitem[Mighell(1999)]{mig1999}
Mighell, K.~J.\ 1999, \apj, 518, 380

\bibitem[Minniti \& Zijlstra(1996)]{min_zij1996}
Minniti, D., \& Zijlstra, A.~A.\ 1996, \apjl, 467, L13

\bibitem[Minniti et al.(1999)Minniti, Zijlstra, \& Alonso]{min_zij_alo1999}
Minniti, D., Zijlstra, A.~A., \& Alonso, M.~V.\ 1999, \aj, 117, 881

\bibitem[Moore et al.(2001)]{moo_etal2001} Moore, B., 
Calc{\'a}neo-Rold{\'a}n, C., Stadel, J., Quinn, T., Lake, G., Ghigna, S., 
\& Governato, F.\ 2001, \prd, 64, 063508 

\bibitem[No{\"e}l \& Gallart(2007)]{noe_gal2007} No{\"e}l, 
N.~E.~D., \& Gallart, C.\ 2007, \apjl, 665, L23 

\bibitem[No{\"e}l et al.(2007)]{noe_etal2007} No{\"e}l, N.~E.~D., 
Gallart, C., Costa, E., \& M{\'e}ndez, R.~A.\ 2007, \aj, 133, 2037 
 
\bibitem[Ortolani \& Gratton(1988)]{ort_gra1988}
Ortolani, S., \& Gratton, R.~G.\ 1988, \pasp, 100, 1405

\bibitem[Peimbert et al.(1994)Peimbert, Colin, \& Sarmiento]{pei_col_sar1994} Peimbert, M., Colin, 
P., \& Sarmiento, A.\ 1994, Violent Star Formation, from 30 Doradus to QSOs, 79 

\bibitem[Pietrinferni et al.(2004)]{pie_etal2004}
Pietrinferni, A., Cassisi, S., Salaris, M., \& Castelli, F.\ 2004, \apj, 612, 168

\bibitem[Purcell et al.(2009)Purcell, Bullock, \& Zentner]{pur_etal2007}
Purcell, C.~W., Bullock, J.~S., Zentner, A.~R.\ 2007, \apj, 666, 20

\bibitem[Sandage(1971)]{san1971} Sandage, A.\ 1971, \apj, 166, 13 

\bibitem[Saviane et al.(2000)]{sav_etal2000} Saviane, I., Rosenberg, A., Piotto, G., \& Aparicio, A.\ 2000, \aap, 355, 966

\bibitem[Schuster \& West(1976)]{sch_wes1976}
Schuster, H.~E., \& West, R.~M.\ 1976, \aap, 49, 129


\bibitem[Seth et al.(2005a)Seth, Dalcanton, \& de Jong]{set_dal_dej2005a} Seth, A.~C., Dalcanton,
J.~J., \& de Jong, R.~S.\ 2005a, \aj, 129, 1331

\bibitem[Seth et al.(2005b)Seth, Dalcanton, \& de Jong]{set_dal_dej2005b} Seth, A.~C., Dalcanton,
J.~J., \& de Jong, R.~S.\ 2005b, \aj, 130, 1574


\bibitem[Smith et al.(1997)Smith, Grillmair, \& Holtzman]{smi_gri_hol1997}
Smith, G., Grillmair, C.~J., \& Holtzman, J.~A.\ 1999, STScI, Proposal 6798 for HST Observations: "The Early Evolution of Local Group Dwarf Irregular Galaxies".

\bibitem[Spergel et al.(2007)]{spergel_etal2007} Spergel, D.~N., et al.\ 2007, \apjs, 170, 377 

\bibitem[Stetson(1994)]{ste1994}
Stetson, P.~B.\ 1994, \pasp, 106, 250

\bibitem[Stinson et al.(2009)]{stin_etal2009}
Stinson, G.~S., Dalcanton, J.~J., Quinn, T., Gogarten, S.~M., Kaufmann, T., Wadsley, J.\ 2009, \mnras, in press

\bibitem[Tikhonov(2005)]{tik2005} 
Tikhonov, N.~A.\ 2005, Astronomy Reports, 49, 501 

\bibitem[Tikhonov(2006a)]{tik2006a} 
Tikhonov, N.~A.\ 2006, Astronomy Letters, 32, 149 

\bibitem[Tikhonov(2006b)]{tik2006b} 
Tikhonov, N.~A.\ 2006, Astronomy Reports, 50, 517 

\bibitem[Tikhonov et al.(2005)Tikhonov, Galazutdinova, \& Drozdovsky]{tik_gal_dro2005} Tikhonov,
N.~A., Galazutdinova, O.~A., \& Drozdovsky, I.~O.\ 2005, \aap, 431, 127

\bibitem[Turner et al.(1998)]{tur1998} Turner, A., et al.\ 
1998, \apj, 505, 207 

\bibitem[Vansevi{\v c}ius et al.(2004)]{van_etal2004} 
Vansevi{\v c}ius, V., et al.\ 2004, \apjl, 611, L93 

\bibitem[van de Rydt et al.(1991)van de Rydt, Demers, \& Kunkel]{ryd_dem_kun1991}
van de Rydt, F., Demers, S., \& Kunkel, W.~E.\ 1991, \aj, 102, 130

\bibitem[Young \& Lo(1997)]{you_lo1997}
Young, L.~M., \& Lo, K.~Y.\ 1997, \apj, 490, 710

\bibitem[Young et al.(2007)]{you_etal2007} Young, L.~M., Skillman, 
E.~D., Weisz, D.~R., \& Dolphin, A.~E.\ 2007, \apj, 659, 331 


\end{thebibliography}
\end{document}